\documentclass[]{gji}

\usepackage{graphicx}
\graphicspath{{../../PaperFigs/}}
\usepackage{caption}
\usepackage{url}
\usepackage{color}
\usepackage{graphicx}
\usepackage[displaymath, mathlines]{lineno}
\usepackage[hidelinks]{hyperref}
\usepackage{amsmath,amssymb}
\usepackage{enumitem}
\usepackage{times}
\usepackage{csquotes}
\usepackage{nicefrac}
\usepackage{multirow}

\newcommand\upi{\mathord{\mathrm{I}}}

\usepackage{mathtools, cuted}

\usepackage{titlesec}
\titleformat{\subsubsection}[runin]
{\normalfont\bfseries}{\thesubsubsection}{1em}{}
\begin{document}

\begin{center}
\textbf{\Large {Elasto-thermo-visco-plastic numerical modelling from a laboratory to geodynamic scale: implications for convergence-driven experiments}} \\[20pt]
\textcolor{blue}{\small Manuscript submitted to Geophys. J. Int.}\\[20pt]

Ekeabino Momoh${^{1,2}}$ Harsha S. Bhat$^1$ and Steve Tait$^{2,3}$

\begin{enumerate}
\small
\it
\itemsep0em
\item{Laboratoire de G\'{e}ologie, \'{E}cole Normale Sup\'{e}rieure, CNRS-UMR 8538, PSL Research University, Paris, France.}
\item{Institut de Physique du Globe de Paris, UMR 7154 Universite de Paris \emph{75005} Paris, France}
\item{Geosciences Environnement Toulouse - GET, Universit\'{e} Toulouse III - Paul Sabatier,  CNRS UMR 5563,\\  \emph{31400} Toulouse, France}
\end{enumerate}

\end{center}
\vspace{0.2cm}

\noindent

\noindent{\bf The development of a subduction zone, whether spontaneous or induced, encompasses a stage of strain localization and is epitomized by the growth of lithospheric-scale shear bands. Our aim in this paper, using a solid-mechanical constitutive description relevant for oceanic lithosphere, is to investigate factors that promote or inhibit localization of deformation in brittle and ductile regimes in convergence-driven numerical experiments. We used the Drucker-Prager yield criterion and a non-associative flow rule, allowing viscoplastic deformation to take directions independent of the preferred direction of yield. We present a step-by-step description of the constitutive law and the consistent algorithmic tangent modulus. The model domain contains an initial weak-zone on which localization can potentially nucleate. In solving the energy conservation problem, we incorporate a \enquote{heat source} term from the mechanical deformations which embodies the irreversible plastic work done. This work term couples the energy equation to the constitutive description, and hence hence the stress balance, via the evolving temperature field. On a sample-scale, we first conduct a series of isothermal benchmark tests.  We then explore behavior including shear heating and volumetric work both separately and in concert. and thereby address the (in)significance  of the latter, and hence assess their potential importance. We find that dilatational effects mostly enhance both shear band development and shear heating. We also observe that high temperature promotes shear band development whereas high confining pressure inhibits it, and infer that the competition between these factors is likely to be the major influence on the position within the lithosphere where shear bands nucleate. Furthermore, we explore the role of size and orientation of the weak zone, thermal diffusion, imposed strain rate and varying rheological parameters on deformation styles. Finally we present the initial steps in an application to a key geodynamic problem, namely the deformation involved in intra-oceanic subduction initiation. We explore at sample scale conditions covering those relevant for the whole lithosphere, and then present preliminary large-scale calculations. We infer that inclined lithospheric-scale shear bands on which subduction can initiate are most likely to nucleate at the base of the lithosphere and propagate upward as heating proceeds.}

\vspace{0.2cm}
\noindent

\section{Introduction}

Subduction (convergent) zones, where mantle lithosphere descends into the Earth's interior, represent some of the sites of the most active tectonic regions in the globe, both from a seismic and volcanic point of view. The formation of a subduction zone involves a stage of strain localization in a compressional tectonic regime during which major plate-boundary faults develop \cite[for e.g.]{lieb1998}, followed by a stage of thermal evolution towards some kind of \enquote{steady-state}. The processes involved in the formation/development of a new subduction zone have not been well understood, nor is the thermal structure explicitly known. A consensus view has developed in the literature that one indispensable ingredient in the physical description of a steady-state subduction zone is a so-called \enquote{corner flow}, whereby the downgoing plate viscously entrains mantle material \cite{mckenzie1969}. Several authors have indeed explored the first order thermal structure that becomes established as heat removal by conduction comes to balance that advected in by the corner flow \cite{england2010} Nevertheless, it has been recently pointed out that the surface heat flux in the ensuing model steady state is much lower than that typically observed at the arc (ref). So while we do not deny the likelihood of a corner flow, we see a need to re-examine the energy budget carefully, Because a corner flow can only be established after a protracted initial period of deformation, it is worth asking what are the potential influences of this stage of initiation. Different approaches in treating the modelling domain lead to different outcomes. Factors like boundary conditions, rheologic description, mineralogic reactions, addition of water and frictional weakening mechanisms can all play a role and their energetic importance needs to be sorted out.

In a bid to analyse this sequence of events and improve understanding of the underlying deformation, we report on model simulations at laboratory scale which can both explicity follow the strain localization process and the thermal evolution of the shear zone and surrounding model. We include specifically a \enquote{source term} that is due to irreversible plastic deformational work, and incorporate a rheologic description appropriate for geologic materials of the upper mantle. The distribution of the heating effect is linked to the deformation field at the scale of the shear band and to the geometry of  localized  deformation or shear zones that can interact. Localized heating can both lead to further localization or to large hot zones where localization is inefficient, because of strong shear in competing directions. These are complex feed-back loops. The current paper aims to benchmark our numerical code as a prelude to carrying out simulations incorporating these complex feed-back loops at larger (geological) scale. 

We first provide a first order description of the interplay between deformation and thermal structure that develops in a geometrically-simple compression experiment, exploring for example, the contributions of shear heating and volumetric heating or cooling (dilatancy-controlled), the significance of dilatancy and plastic versus viscoplastic deformation; the role of thermal diffusivity and the initiation angle of an initial weak zone. Furthermore, we investigate the influence of loading rates on the deformation. Our use of this terminology shall be clarified below in the context of a specific rheologic description. 

The material properties utilized for the compression tests to study viscoplastic deformation have been sourced from the experimental rock mechanics literature, and are those typically used in geodynamic models (references provided below) and the initial thermal conditions and accompanying lithostatic stress state are similarly estimated. The strategy is to provide a chart for the competing conditions that help to localize (or delocalize) deformation, and above all to investigate the potential thermal feedbacks between rheology, deformation and temperature field which are coupled via the energy equation. 

We begin by stating the conservation laws and then give a step by step description of the constitutive laws and solution methods we have developed. Thereafter, we present the results of some benchmark tests on plasticity following earlier numerical studies by other authors, and we present the results of the exploration of the different contributions in terms of material properties, geometrical controls, thermal state and loading conditions.  
\section{Theoretical framework}
\subsection{Conservation laws}
Our starting point is the balance of linear momentum given by:
\begin{equation}
\dfrac{\partial{{\sigma}_{ij}}}{\partial{x}_j} +f_i=\rho\dfrac{\partial^2{u_i}}{\partial{t}^2},
\end{equation}
\noindent where ${\sigma_{ij}}$ represents components of the Cauchy stress tensor, ${f_i}$ represent internal body forces, in this case the  lithostatic stress state due to the weight of the material given by ${\rho g z}$, with ${\rho}$ as density, ${g}$ as gravitational acceleration, and ${z}$ as depth (pointing downwards); ${x}$  represents spatial variables in the Cartesian coordinates; ${{u_i}}$ are the components of displacement. Deformational heating arises from irreversible thermomechanical work, i.e., due to creep and plastic deformation. We can account for heat generation and transport in the conservation of energy which reads:
\begin{equation} 
\dfrac{\partial{T}}{\partial{t}}=\alpha_\textrm{th}{\nabla^2 T}+\beta\dfrac{{{{\sigma_{ij}}\dot\varepsilon}^\textrm{vp}_{ij}}}{\rho C_p},
\label{eqn:thermal}
\end{equation} 
where ${C_p\;(J.Kg^{-1}.K^{-1})}$ is the specific heat capacity and $\alpha_\textrm{th}\;(m^{2}s^{-1})$ is the thermal diffusivity. ${{{\dot\varepsilon}}^{\textrm{vp}}_{ij}}$ represents the deformational contribution to strain rate from creep and plastic deformations, $\sigma_{ij}\dot{\varepsilon}^\textrm{vp}_{ij}$ is the contribution to heating from deformational work \cite{ravi} with ${\beta}$, the Taylor-Quinney coefficient which quantifies the proportion of deformational work which is converted to heat. 

\subsection{Constitutive framework}
Our constitutive law incorporates elastic strain, ductile creep and plastic flow laws with inelastic work being converted to heat.  

Non-Associative viscoplasticity is formulated using the Drucker-Prager failure criterion. We summarize this rheologic description as elasto-thermo-visco-plastic.

\noindent Our formulation relies on small strain assumption at each iterative step of the numerical calculation. In our approach, we compute the stress evolution, creep and plastic flow relying on an additive decomposition of the total strain tensor into an elastic and viscoplastic strains, represented by superscripts ${e}$ and ${vp}$ below:
\begin{equation}
{\varepsilon}_{ij} = {\varepsilon}^\textrm{e}_{ij}+{\varepsilon}^\textrm{vp}_{ij}.
\end{equation}
\subsubsection{Elastic Rheology}
The elastic part of material response is treated as an isotropic solid characterized by its Young's modulus, $E$, and Poisson's ration, $\nu$. 
\subsubsection{Yield Criterion} 
Whether or not plastic yielding occurs is assessed using the pressure-sensitive Drucker-Prager criterion given by \cite{dpg1952}:
\begin{equation} 
\Phi_{Y}({\sigma}_{ij}, c) = \sqrt{J_\textrm{II}}+\alpha_1P-\alpha_2c \ge 0.
\end{equation}
\noindent Where $J_\textrm{II}=s_{ij}s_{ij}/2$ represents the second invariant of the deviatoric stress tensor, $\left({s_{ij}=\sigma_{ij}-\sigma_{kk}\delta_{ij}/3}\right)$, ${P=I_1/3=\sigma_{kk}/3}$ is related to the the first invariant of the stress tensor, ${c}$ is the cohesion which may depend on the viscoplastic strain history; ${\alpha_1}$ and ${\alpha_2}$ are material-dependent constants which are given functions of the internal friction angle (${\varphi}$) as follows:
\begin{equation} 
\alpha_1=\dfrac{3 \tan{\varphi}}{\sqrt{9+12\tan{^2\varphi}}}, \alpha_2=\dfrac{3}{\sqrt{9+12\tan{^2\varphi}}}.
\end{equation}
\subsubsection{Flow Law}
We write the law to describe flow once viscoplastic deformation has begun following the standard formulation \cite{desai1987,vermeer1990}:
\begin{equation}
\dot{\varepsilon}_{ij}^\textrm{vp} = \dot{\gamma}({\sigma_{ij}},T)\dfrac{\partial{\Phi_{F}}}{\partial{\sigma}_{ij}}
\label{viscoflow}
\end{equation}
specifying a magnitude and direction of viscoplastic flow, with ${\dot{\gamma}}$ as a non-negative quantity defined as the viscoplastic consistency parameter describing the magnitude of viscoplastic flow \cite{simo1998}; $\Phi_{F} = \sqrt{J_\textrm{II}} + \alpha_3 P$ is the viscoplastic flow potential and $\alpha_3=3 \tan{\psi}/ \sqrt{9+12\tan{^2}\psi}$, ${\psi}$ is the dilatancy angle whose value may associate or dissociate the viscoplastic flow potential from the yield criterion \cite{vermeer1984}. The derivative of the viscoplastic flow potential with respect to the stress tensor describes the direction of viscoplastic flow: 
\begin{equation}
 \dfrac{\partial{\Phi_{F}}}{\partial{\sigma}_{ij}}=\dfrac{{s}_{ij}}{2\sqrt{J_\textrm{II}}}+ \dfrac{\alpha_3}{3}\delta_{ij}.
\end{equation}
The equation above indicates that there is a deviatoric and volumetric (dilatant) contribution to the direction of viscoplastic flow. Non-zero dilatancy removes the assumption of an incompressible material, i.e., one for which pressure changes due to viscoplastic deformation do not result in a net volume change in the material \cite{turcotte2002}.

The viscoplastic consistency parameter $\dot{\gamma}$ assumes various functional forms depending on the specific problem and material \cite{neto}, we utilize the functional form of \cite{boyle},
\begin{equation}
{\dot{\gamma}({\sigma_{ij}},T)=\left<{\Phi_{Y}}\right>^mf(T)}
\label{functionals}
\end{equation}
where ${\left<\cdot\right>}$ is the Macaulay bracket. $m$ represents the power law exponent that describes the sensitivity to stress during viscoplasticity. The temperature dependence term is given by, 
\begin{equation}
f(T)=A{\textrm e}^{-\frac{E_a}{RT}}
\label{disloc}
\end{equation}
Here $E_a$ is the activation energy, $R$ is the molecular gas constant, $T$ is the absolute temperature (in degrees Kelvin, $K$) and $A$ is the pre-exponential factor. This accounts for creep by assuming the dominant creep mechanism is dislocation creep given by the Norton-Hoff constitutive equation for steady-state creep \cite{skrzypek,kohlstedt,ranalli}.
\subsubsection{Hardening laws}
We incorporate two hardening laws in our constitutive formulation. First is a linear cohesion hardening which depends on the deformation history and second is friction hardening. The cohesion hardening is given by:
\begin{equation}                                           
c = c_0 + H{\bar{\varepsilon}},
\end{equation}
\noindent where ${c}$ is the updated cohesion, ${c_0}$ is the initial cohesion and ${\bar{\varepsilon}}$ is a quantity that records the deformation history during viscoplastic deformation:
\begin{equation}
\dot{\bar{\varepsilon}}=\alpha_2\dot{{\gamma}}
\end{equation}
The friction hardening follows the form proposed in finite element models of strain localization in pressure-dependent materials that are undergoing multi-axial loading \cite{leroy1989}:
\begin{equation}
\sin \varphi=\sin{\varphi}_{i}+\dfrac{2\left(\sin{\varphi_f}-\sin{\varphi_i}\right)\sqrt{\varepsilon_{\textrm{eff}}^\textrm{vp}\varepsilon_\textrm{crit}^\textrm{vp}}}{\varepsilon_{\textrm{eff}}^\textrm{vp}+\varepsilon_{\textrm{crit}}^\textrm{vp}}.
\end{equation}
\noindent Where ${\varphi_i}$ is the initial friction angle, ${\varphi_f}$ is the maximum friction angle reached when the effective viscoplastic strain invariant ${(\varepsilon_{\textrm{eff}}^\textrm{vp})}$ reaches a critical value ${(\varepsilon_\textrm{crit}^\textrm{vp})}$. The accumulated viscoplastic strain (magnitude of deformation) is given by: 
\begin{equation}
{\varepsilon}_{\textrm{eff}}^\textrm{vp} = \sqrt{\dfrac{2}{3}{\varepsilon}_{ij}^\textrm{vp}{\varepsilon}_{ij}^\textrm{vp}}
\label{effective_viscoplastic_strain_inv} 
\end{equation}
\section{Numerical Framework}
The above formulations to solve the mechanical and thermal diffusion parts of the system have been coded into UMAT and UMATHT subroutines in Abaqus \cite{abaqustheory,abaqus}. UMAT provides the user with the total strain increment at a given time step and requires the user to compute the corresponding stress increments and the Consistent Algorithmic Tangent Modulus. Below we describe the details.
\subsubsection{Stress Update Scheme}
Most commonly, visco-elasto-plastic models compute viscous and plastic deformation terms separately. In practice, a correction to deviatoric stress due to dislocation or diffusion creep is computed, and the corrected stress then used to assess plastic yielding (\cite{jacquey,zhou2021}). This approach introduces an, arguably artificial, \enquote{plastic viscosity} term whose value is not usually well-constrained.  In contrast, as laid out below, we combine creep and plastic deformation into a single computational step as well as incorporating frictional and cohesion hardening.

At each time step we first assume that the strain increment, ${{\Delta{\varepsilon}^\textrm{e}_{ij}}}$, at that time step is fully elastic. We then compute a trial stress.
\begin{equation}
{\sigma}^{\textrm{trial}}_{ij} = C^\textrm{e}_{ijkl}\left({\varepsilon}^\textrm{e}_{kl}+{\Delta\varepsilon}^\textrm{e}_{kl}\right)
\end{equation}
If the yield criterion, $\Phi_{Y}$,  based on the trial stress, is violated then we need to compute the viscoelastic strain increment. This requires us to compute the plastic multiplier increment, $\Delta\gamma$, at the current time step. Once we obtain $\Delta\gamma$ then,
\begin{equation}
\Delta{\varepsilon}^\textrm{vp}_{ij}  = \Delta{\gamma}\left(    \dfrac{{s}^{\textrm{trial}}_{ij}}{2\sqrt{J_\textrm{II}({s}^{\textrm{trial}}_{ij})}}+ \dfrac{\alpha_3}{3}\delta_{ij}\right)
\end{equation}
\begin{equation}
{{{s}_{ij} ={\left(1-\dfrac{G\Delta{\gamma}}{\sqrt{J_\textrm{II}({s}^{\textrm{trial}}_{ij})}}\right)}{s}_{ij}^{\textrm{trial}}}}
\label{deviatoric_update}
\end{equation}
\begin{equation}
{ P=P^{\textrm{trial}}-{\alpha_3}\Delta{\gamma}K}
\label{pressure_update}
\end{equation}
where $K$ is the Bulk modulus.
\noindent The corresponding increment in temperature is given by:
\begin{equation}
\Delta{T}=\dfrac{\sigma_{ij}\Delta{\varepsilon_{ij}^\textrm{vp}}}{\rho C_p}.
\label{temperaturechange}
\end{equation}
\noindent Where the numerator in Equation \ref{temperaturechange} can be treated as a \enquote{heat source term}, i.e., the conversion of deformational work to heat \cite{ponthot2004}. To find the unknown ${\Delta{\gamma}}$, we proceed by defining a viscoplastic return mapping function following the formalism of \cite{net}:
\begin{equation}
\widetilde{\Phi}(\Delta\gamma) \equiv {\Phi^m_{Y}(\Delta\gamma)}f(T) - \dfrac{\Delta\gamma}{\Delta t}
\label{returnmap}
\end{equation}
where $\Delta t$ is the current time increment. Equation \ref{returnmap} is a non-linear algebraic equation which has no analytical solution. We therefore utilize an iterative search for the solution. We utilize a Newton-Raphson scheme \cite{yqma} to solve $\widetilde{\Phi}(\Delta\gamma) = 0$. The derivative of the non-linear return mapping equation with respect to the unknown, $d\widetilde{\Phi}/d\Delta\gamma$, is required by the Newton-Raphson scheme to find the unknown. This derivative is given by:
\begin{equation}
\dfrac{d\widetilde{\Phi}}{d\Delta\gamma} = -c_{1}m\Delta{t}f(T)(\Phi_{Y}-c_{1}\Delta\gamma)^{m-1} - 1
\end{equation}
where $c_{1} = \left(G+\alpha_1\alpha_3K+\alpha_2^2H\right).$
The introduction of invariants reduces the system of equations depending on the number of tensor components (for plain stress, plain strain, axisymmetric or 3-D problems) to a single equation. The only unknown in this equation and consequently the update scheme as shown above, is the (visco)plastic multiplier. The goal of solving the viscoplastic return mapping equation is to compute this unknown using the Newton-Raphson scheme.
\subsubsection{Consistent Algorithmic Tangent Modulus (CATM)}
Finally, using the Newton scheme to solve the global equations, an important consideration is the convergence of the algorithm. To this end, we estimate a convergence matrix obtained by linearizing the stress update scheme given in the equations above. To this end, we estimate a convergence matrix obtained by linearizing the stress update schemes for deviatoric stresses and pressure given in Equations \ref{deviatoric_update} and \ref{pressure_update}. We proceed by taking small increments in the deviatoric stress and pressure with respect to the trial elastic state. The scheme for estimating the convergence matrix is obtained from:
\begin{equation}
C_{ijkl}^\textrm{vp}={\dfrac{d{\sigma}_{ij}}{d\varepsilon_{kl}^{\textrm{e}\;\textrm{trial}}}=\dfrac{d{s}_{ij}}{d\varepsilon_{kl}^{\textrm{e}\;\textrm{trial}}}+\delta_{ij}\left(\dfrac{dP}{d\varepsilon_{kl}^{\textrm{e}\;\textrm{trial}}}\right)}
\end{equation}
The expression for deviatoric stresses can be equivalently written in terms of the deviatoric strains (${{e}_{ij}^{\textrm{trial}}}$) taking account that ${s_{ij}^\textrm{trial}=2G{e}_{ij}^{\textrm{trial}}}$, and ${{e}_{ij}^{\textrm{trial}}}={{\varepsilon}_{ij}^{\textrm{trial}}}-{{\varepsilon}_{kk}^{\textrm{trial}}}/3$:
\begin{equation}
{{{s}_{ij} =2G{\left(1-\dfrac{G\Delta{\gamma}}{\sqrt{J_\textrm{II}({s}^{\textrm{trial}}_{ij})}}\right)}{e}_{ij}^{\textrm{trial}}}}.
\end{equation}
We define some quantities to be used in the expression for the consistent Jacobian:
\begin{equation}
B=
\\
\\
-m\Delta{t}f(T)\left({\dfrac{d\widetilde{\Phi}}{d\Delta\gamma}}\right)^{-1}{\left(\Phi_Y-c_1\Delta{\gamma}\right)^{m-1}}
\end{equation}
The elements for assembling the consistent Jacobian matrix are given by derivatives of the stress states: 
\begin{eqnarray}
\dfrac{d{s}_{ij}}{d\varepsilon_{kl}^{\textrm{e}\;\textrm{trial}}}  &= 2G{\left(1-\dfrac{G\Delta{\gamma}}{\sqrt{J_\textrm{II}({s}^{\textrm{trial}}_{ij})}}\right)}\mitbf{\upi}^{\textrm {dev}}_{ijkl}\nonumber\\&+2G\left(\dfrac{\Delta{\gamma}}{\sqrt{2}||e_{ij}^{\textrm{e}\;\textrm{trial}}||}-GB\right)\dfrac{{e}_{ij}^{\textrm{trial}}}{{e}_{\textrm{norm}}^{\textrm{trial}}}\dfrac{{e}_{kl}^{\textrm{trial}}}{{e}_{\textrm{norm}}^{\textrm{trial}}}\nonumber\\&-\sqrt{2}\alpha_1GKB\dfrac{{e}_{ij}^{\textrm{trial}}}{{e}_{\textrm{norm}}^{\textrm{trial}}}\dfrac{{e}_{kl}^{\textrm{trial}}}{{e}_{\textrm{norm}}^{\textrm{trial}}}\delta_{kl},
\label{deviatoric contribution}
\end{eqnarray}
{\centering{
\begin{eqnarray}
\delta_{ij}\dfrac{dP_{n+1}}{d\varepsilon_{kl}^{\textrm{e}\;\textrm{trial}}}&=K\left(1-\alpha_1\alpha_3KB\right)\delta_{ij}\delta_{kl}\nonumber\\ & -\sqrt{2}\alpha_3GKB\delta_{ij}\dfrac{{e}_{kl}^{\textrm{trial}}}{{e}_{\textrm{norm}}^{\textrm{trial}}}.
\label{volumetric contribution}
\end{eqnarray}}}
\noindent The final expression for the consistent Jacobian is given by a combination of Equations {\ref{deviatoric contribution}} and \ref{volumetric contribution}:
{\begin{center}
\begin{eqnarray}
C_{ijkl}^\textrm{vp} =2G{\left(1-\dfrac{G\Delta{\gamma}}{\sqrt{J_\textrm{II}({s}^{\textrm{\textrm trial}}_{ij})}}\right)}\mitbf{\upi}^{\textrm {dev}}_{ijkl} \nonumber
\\+2G\left(\dfrac{\Delta{\gamma}}{\sqrt{2}{e}_{\textrm{norm}}^{\textrm{trial}}}-GB\right)\dfrac{{e}_{ij}^{\textrm{trial}}}{{e}_{\textrm{norm}}^{\textrm{trial}}}\dfrac{{e}_{kl}^{\textrm{trial}}}{{e}_{\textrm{norm}}^{\textrm{trial}}}
\nonumber\\+K\left(1-\alpha_1\alpha_3KB\right)\delta_{ij}\delta_{kl}\nonumber\\-\sqrt{2}GKB\left(\alpha_1 \dfrac{{e}_{ij}^{\textrm{trial}}}{{e}_{\textrm{norm}}^{\textrm{trial}}}\delta_{kl}+\alpha_3\delta_{ij}\dfrac{{e}_{kl}^{\textrm{trial}}}{{e}_{\textrm{norm}}^{\textrm{trial}}}\right)
\end{eqnarray}\end{center}}
\noindent Where ${{e}_{\textrm{norm}}^{\textrm{trial}}={||{e}_{ij}^{\textrm{e}\;\textrm{trial}}||}}$, ${\mitbf{\upi}_{ijkl}^{\textrm{dev}}  = \left(\delta_{ik}\delta_{jl} + \delta_{il}\delta_{jk}\right)/2-\delta_{ij}\delta_{kl}/3}$ is the fourth order deviatoric projection tensor, which extracts the deviatoric components of a stress or strain tensor.
\subsubsection{Return mapping to apex of Drucker-Prager cone}
The above theoretical formulation and numerical implementation maps the stress state to the smooth part of the Drucker-Prager cone. Consistency checks for the validity of return mapping are carried out via:
\begin{equation}
\sqrt{J_\textrm{II}({s}^{\textrm{trial}}_{ij})}-{G\Delta{\gamma}}.
\end{equation}
\noindent
If the above quantity is negative, it is appropriate to solve the return mapping equation to the apex of the Drucker-Prager cone instead, where deviatoric stresses disappear. We therefore define a new plastic consistency condition corresponding to return mapping to the apex and cast the viscoplastic constitutive equation in the following framework:
\begin{equation}
\widetilde{\Phi}\left(\Delta\gamma_{\textrm{apex}}\right)={{f(T)^{\nicefrac{1}{m}}}\left(\bar{\beta}c\left(\bar{\varepsilon}\right)-P\right)}-\Delta\gamma_\textrm{apex}^{\nicefrac{1}{m}},
\label{returnmap1}
\end{equation}
\noindent
where,
\begin{equation}
c=c_0+H(\bar\varepsilon_\textrm{apex})
\end{equation}
\begin{equation}
{\bar\varepsilon}_\textrm{apex}=\int_0^t\bar\alpha\dot{\gamma}_{\textrm{apex}} dt
\end{equation}
and
${\bar\alpha=\dfrac{\alpha_2}{\alpha_1}}$, ${\bar\beta=\dfrac{\alpha_2}{\alpha_3}}$. After finding the zeros of the residual viscoplastic equation using the new consistency condition (we adopt the Newton-Raphson approach), we update the stress state as follows:
\begin{equation}
P=P^{\textrm{trial}}-K\Delta\gamma_\textrm{apex}.
\label{apexequationderivative}
\end{equation}
\begin{equation}
{\sigma}_{ij}=P\delta_{ij}.
\label{pressureupdate}
\end{equation}
To assemble the consistent algorithmic tangent modulus, we proceed in a similar fashion with the smooth return by taking small variations in Equation \ref{apexequationderivative}
\noindent The final expression for the consistent algorithmic tangent modulus for the return mapping to the apex of the Drucker-Prager cone is given by:
\begin{equation}
{\dfrac{d{\sigma}_{ij}}{d\varepsilon_{kl}^{\textrm{e}\;\textrm{trial}}} =\delta{ij}\left(\dfrac{dP}{d\varepsilon_{kl}^{\textrm{e}\;\textrm{trial}}}\right) 
=K\left(1-BK\right)\delta_{ij}\delta_{kl}}.
\end{equation}
\noindent where
\begin{equation}
B=\left(\dfrac{\left(\Delta{t}f(T)\right)^{\nicefrac{1}{m}}}{{\left(\bar\alpha \bar\beta H+K\right)\left(\Delta{t}f(T)\right)^{\nicefrac{1}{m}}-\dfrac{1}{m}{\left(\Delta{\gamma_\textrm{{apex}}}\right)^{{{\nicefrac{(1-m)}{m}}}}}}}
\right).
\end{equation}
\subsubsection{Thermal balance}
In solving the thermal diffusion problem, the finite element solver requires a heat generation term, \enquote{R\textsubscript{PL}}, as heat source (plastic heat generation rate). Specification of this term is done in the stress balance routine. The next step in the thermal subroutine is to account for the  change in the internal energy of the system, which can be identified in Equation \ref{eqn:thermal}: 
\begin{equation} 
\dot{U}=C_p\dfrac{\partial T}{\partial t},
\end{equation} 
and can be approximated in incremental form (using a finite difference approximation) as:
\begin{equation} 
\Delta{U}=C_p\Delta{T},
\end{equation} 
The elements summarized above are incorporated into the Abaqus heat transfer user subroutine (UMATHT) which solves the heat transfer/conversion problem and thermal diffusion after the mechanical deformation. As specified in Equation {\ref{temperaturechange}}, we use the thermo-mechanical deformations as input in the \enquote{heat generation term (RPL)} variable as the heat source Abaqus requires to solve the thermal balance equation with the thermal subroutine (UMATHT). This is the traditional source of local deformational heating in thermo-mechanical deformation problems \cite{ostwald19}. The remaining elements utilized in the UMATHT include variation of internal energy with respect to temperature and spatial gradients of temperature in order to compute the diffusive thermal flux and the variation with spatial gradients. 
\section{Algorithmic testing and validation}
\subsection{Strain localization in brittle and ductile regimes}
\subsubsection{Benchmark tests for plasticity}
Shear bands are not typically observed in model studies when the sample is described by homogeneous material properties; therefore localizing deformation requires an impurity in the form of differences in material properties \cite{kaus2010}, deformation-related grain-size reduction \cite{thielmann2015}, shear heating \cite{thielmann2012,willis2019}, material anisotropy \cite{pardoen15} or damage rheology. We test the robustness of our constitutive update on localization studies in pressure-dependent materials \cite{leroy1989,ortiz1990}. This is one for which the strain components with direction normal to the plane of deformation is not considered \cite{zienctay}.

\begin{figure}
\centering{
\includegraphics[width=\columnwidth]{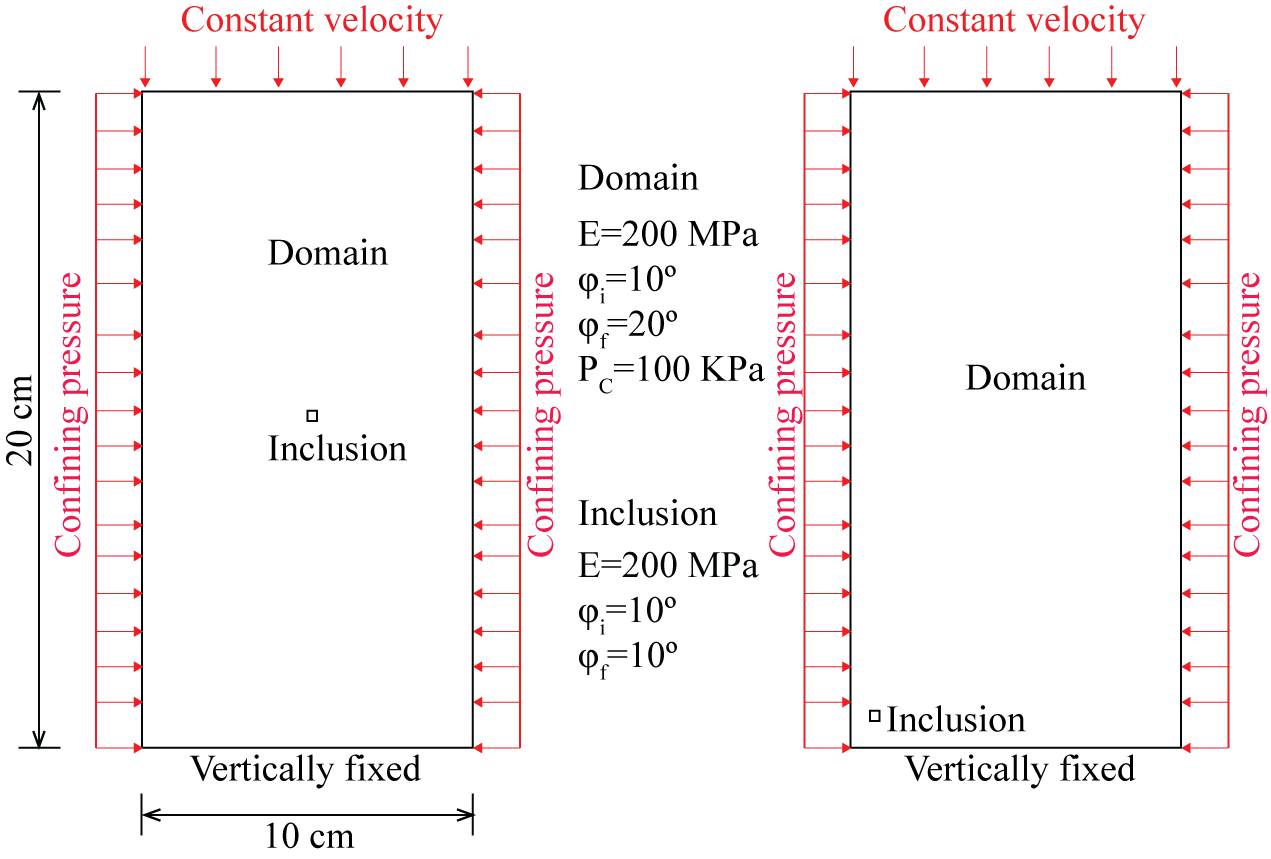} }
\caption{Model setup and boundary conditions for benchmark example of shear band localization in plasticity. The model is homogeneous except for a single element heterogeneity which differs in its friction properties.}
\label{shearbandsetup}
\end{figure}

\noindent Two cases investigated in these biaxial compression tests include: (1) a heterogeneity located at the center of a model, and (2) a heterogeneity at the bottom left of the model (Figure {\ref{shearbandsetup}}). The same material properties utilized by  \cite{leroy1989,ortiz1990} have been chosen here, with the heterogeneity having different friction and dilatancy angles. The compression experiments are first done without considering the thermal part of the constitutive law and creeping effects, i.e., assuming rate-independent plasticity. Confining pressures were applied on the right and left boundaries which are kept constant throughout the experiment; the bottom boundary was fixed for vertical motion while a constant velocity was applied at the top boundary. Four-node quadilateral elements have been adopted with four integration points to compute the displacements and resultant strain increments from which the other quantities can be computed.  Once the Drucker-Prager yield criterion is violated, we compute plastic flow as described above.

\begin{figure}
\centering{
\includegraphics[width=6cm]{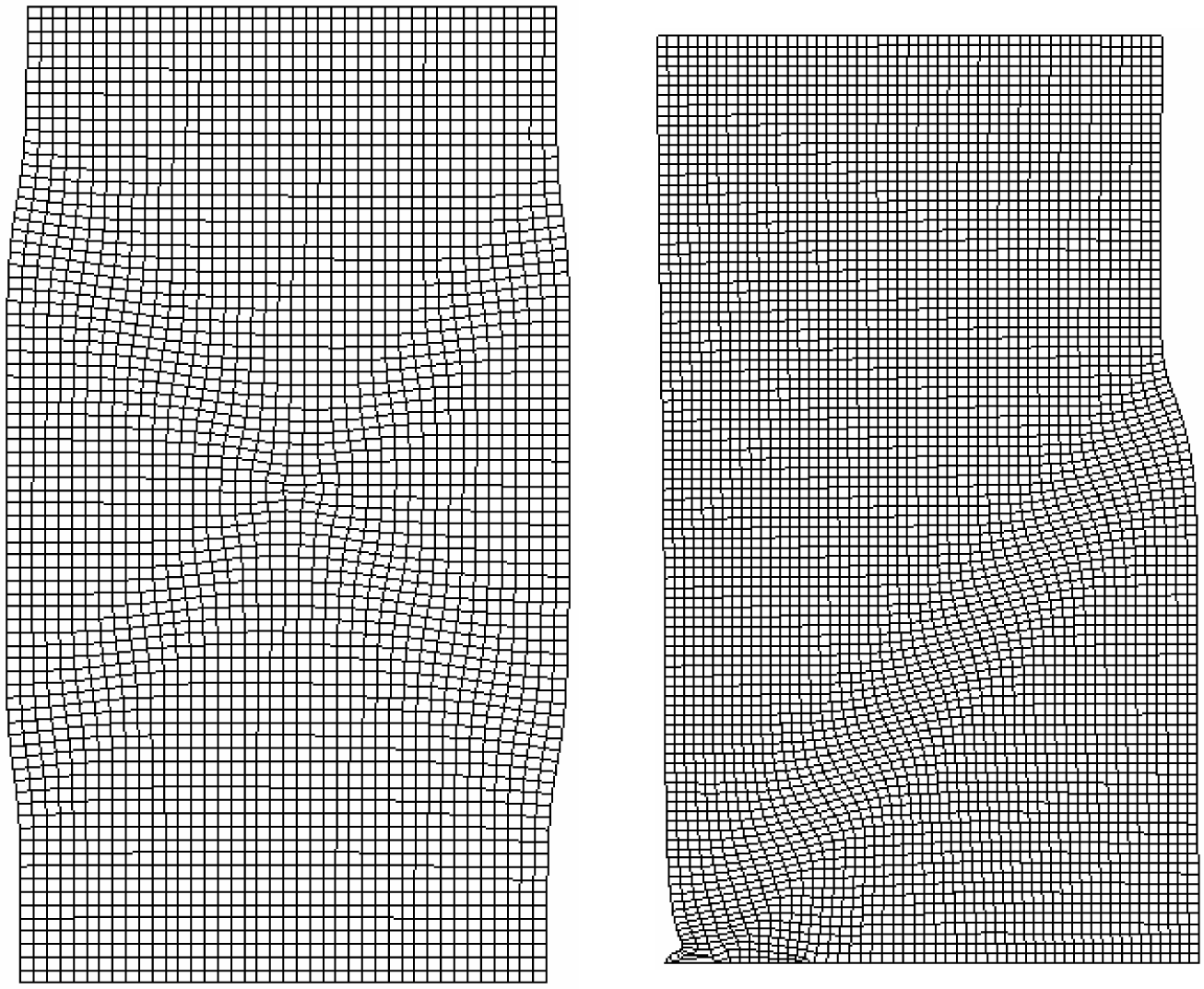}}
\caption{Examples of deformed meshes for the case of frictionally weak element located at the center of the model (left) and at the bottom left of the model (right), respectively. The setup is shown in Figure \ref{shearbandsetup}. The deformation has not been exaggerated.}
\centering{
\includegraphics[width=5cm]{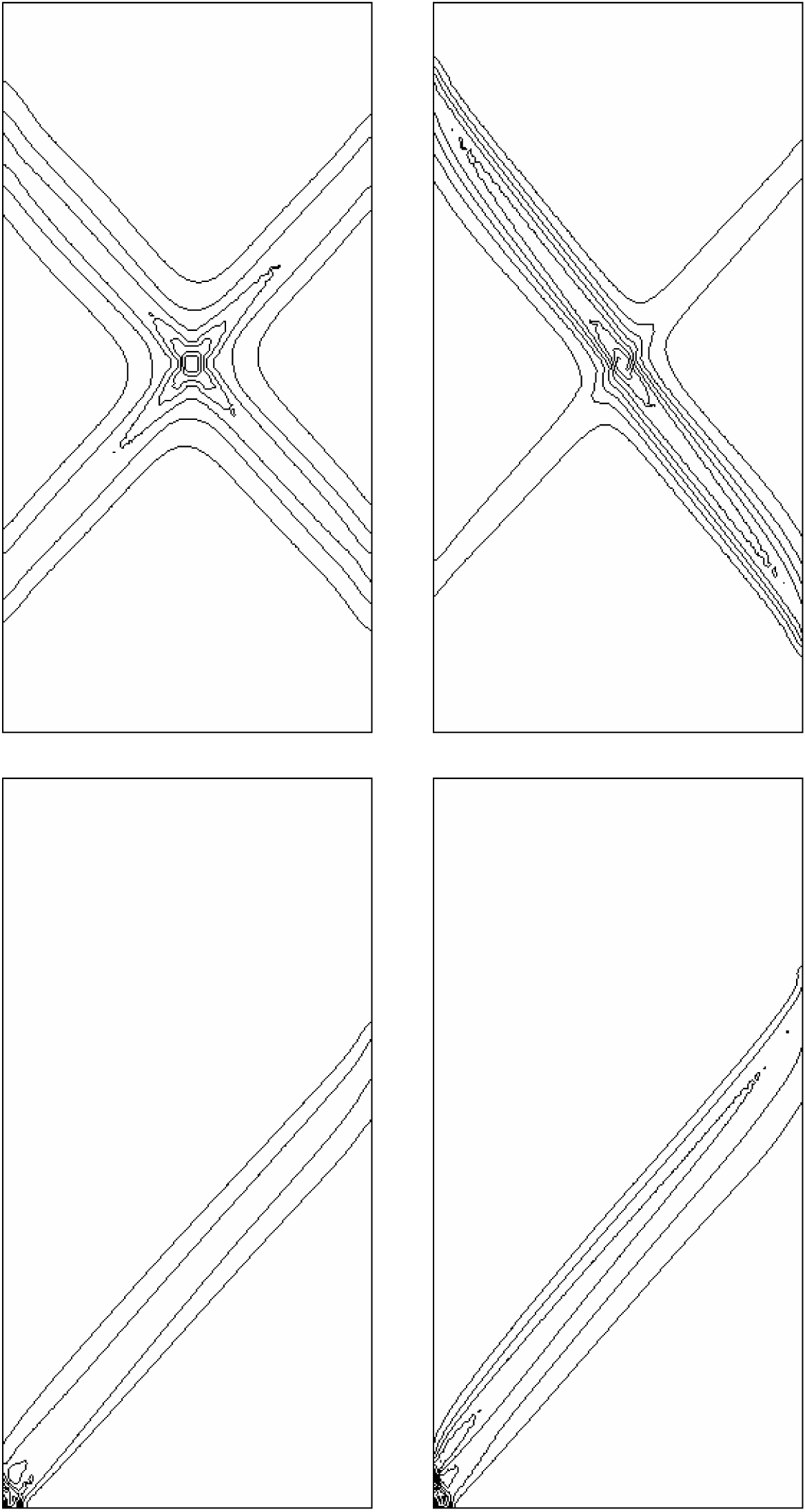}}
\caption{Two examples of contours of effective plastic strain invariant at an intermediate time step and at the end of the prescribed time step. The contours range between 0.37${\%}$ and 143${\%}$ for the examples shown for the heterogeneity at the middle of the model, and range between 0.18${\%}$ and 217${\%}$ where the heterogeneity is located at the bottom left of the model. The maximum contour during the intermediate stage is 61${\%}$ and 1${\%}$ respectively. The setup is shown in Figure \ref{shearbandsetup}. The figure is shown in undeformed configuration.}
\label{deformedmeshandcontour}
\end{figure}

\begin{figure}
\centering{\includegraphics[width=8cm]{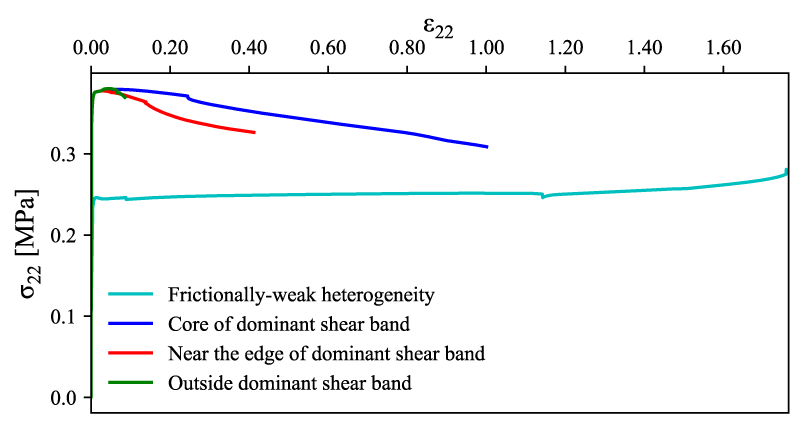}}
\caption{Axial stress-strain curves for integration point elements within the dominant shear band, near the edge of the dominant shear band and outside the dominant shear band for the case an heterogeneity in the middle of the element.}
\label{onedplots}
\end{figure}

The deformed meshes for compression of the two case studies are shown in Figure \ref{deformedmeshandcontour} indicating developing shear bands and consequent deformed shape of the model. Contours of effective plastic strain (Equation \ref{effective_viscoplastic_strain_inv}) at an intermediate and the at the final time step used for this test are shown for the two cases in Figure \ref{deformedmeshandcontour} illustrating the onset of a symmetric pattern of shear band and the fate at the end of the time step with one dominant band localizing the deformation. Despite the prescription of friction and cohesion hardening in the constitutive description, the onset and growth of shear bands and subsequent localization of deformation leads to progressive mechanical softening of the most deformed parts of the material (Figure \ref{onedplots}), but at a given strain, softening in the heart of the shear band is less than in the material just outside. It is also apparent from Figure \ref{onedplots} that the elastic limits are coincident for three cases outside the weak heterogeneity, but the accumulation of deformation are dissimilar; however, hardening is apparent in the weak inclusion, with local softening at some strains. The localization that is expressed by the developing shear bands, thus results from the interaction of the deformation associated with the heterogeneity with the boundary conditions applied. 
\subsubsection{Loading conditions and the role of material properties for shear localization in visco-plasticity}
Since temperature effects have not been taken into consideration in this first plastic localization experiment, one observation is that a weak plastic zone, plastic yielding and non-associative plastic flow rule seem to favour plastic strain localization at this scale. Below, we progressively introduce temperature and creeping effects, i.e., using the full constitutive law to investigate \enquote{ductile} shear localization.    

We utilize the same set-up of the heterogeneity located at the center of the model (Figure \ref{shearbandsetup}) with various material parameters and loading conditions to investigate their effects on the deformation. A constant, but high temperature of 1245 K has been used for these simulations, where creeping effects are dominant. The models are described as M\textsubscript{1} to M\textsubscript{9}, and the corresponding materials properties and loading conditions are shown in  Table \ref{m1tom10}. The deformation pattern mostly initiated as symmetric \enquote{X-shaped} shear bands which apparently influence deformation at the boundaries of the model with the subsequent interaction leading to further deformation. The observed shear band angles range between 58${^\circ}$ and 71${^\circ}$, with varying widths, amplitudes and geometry. Although the shear bands initialized at different angles, the final angles may have resulted from the interactions between the shear bands developing upwards or downwards and the deformation from the boundaries developing in the opposite direction. In some cases, the deformation overprinted the initial shear bands. The shear band angles are measured along straight paths at the core of the initial bands where they are visible at the end of the deformation with respect to the horizontal for the case of vertical compression which we report here.

The sensitivity of the Drucker Prager plastic yield criterion to the value of the stress exponent ${m}$ can be assessed by comparing models M\textsubscript{1} and M\textsubscript{2}. The latter, with a higher exponent, has a comparatively less-developed (i.e., weaker amplitude) shear band (note the scale differences between the M\textsubscript{1} and M\textsubscript{2}). Therefore, when sensitivity to plasticity is considered, it is apparent that localization is  less efficient. Higher saturation friction angles and stress exponent for the weak element and higher friction angle for the rest of the model exhibits slightly less deformation were also explored by comparing the results of models M\textsubscript{2} and M\textsubscript{3}.   
{\small{\begin{table}
\caption{Biaxial compression tests to benchmark numerical implementation of the stress update strategy with focus on plasticity. Applied strain rate applied is 5${\times10^{-16} s^{-1}}$. The test is temperature-independent. Common material properties are: {E} - Young's modulus, ${\nu}$ - Poisson ratio = 0.24, ${\varepsilon^\textrm{vp}_c}$ -  critical strain value = 0.25${\%}$, this represents the strain value for which the friction angle, ${\phi}$, takes the final value ${\phi_f}$ \textit{H} -  hardening modulus = 0 MPa, ${c_0}$ - initial cohesion = 0.02 MPa for the element and 200 MPa for the rest of the model. P\textsubscript{c} represents confining pressure.}
\label{m1tom10}
\begin{center}
\begin{tabular}{@{}lccccccc}
\hline
Model & {E} (GPa) & P\textsubscript{c} (MPa)& ${\varphi_i}$ & ${\varphi_f}$ & ${\psi}$ & ${m}$ \\
\hline
M\textsubscript{1} &   & & &  &  & &   \\
\hspace{2mm} Model &185 &0.1 &10 &20 &2 &2  \\
\hspace{2mm} Element&185 &0.1 &10 &10 &2 &2  \\
\hline
M\textsubscript{2} & & & &  &  & &   \\
\hspace{2mm} Model&185  &0.1 &10 &20 &5 &3.5  \\
\hspace{2mm} Element&185 &0.1 &10 &10 &5 &3.5 \\
\hline
M\textsubscript{3} & & & &  &  & &  \\
\hspace{2mm} Model&185  &0.1 &10 &25 &5 &3.5\\
\hspace{2mm} Element&185 &0.1 &10 &17 &5 &4.94\\
\hline
M\textsubscript{4} & & & &  &  & &  \\
\hspace{2mm} Model&185  &80 &10 &25 &5 &3.5 \\
\hspace{2mm} Element&185 &80 &10 &17 &5 &4.94 \\
\hline
M\textsubscript{5} & & & &  &  & &  \\
\hspace{2mm} Model&185  &200 &10 &25 &5 & 3.5 \\
\hspace{2mm} Element&185 &200 &10 &17 & 5 &4.94  \\
\hline
M\textsubscript{6} & &  & &  &  & &  \\
\hspace{2mm} Model&185  &200 &10 &25 &8 &3.5\\
\hspace{2mm} Element&185 &200 &10 &17 &8 &4.94 \\
\hline
M\textsubscript{7} & & & &  &  & &  \\
\hspace{2mm} Model&185  &200 &10 &25 &8 &3.5  \\
\hspace{2mm} Element&185 &200 &10 &17 &10 &4.94 \\
\hline
M\textsubscript{8} & & & &  &  & &\\
\hspace{2mm} Model &185 &200 &23.6 &36.9 &10 & 3.5  \\
\hspace{2mm} Element &185&200 &23.6 &30 &10 &4.94\\
\hline
M\textsubscript{9} & &  & &  &  & & \\
\hspace{2mm} Model & 185 &200 &23.6 &36.9 &10 & 3.5  \\
\hspace{2mm} Element& 100 &200 &23.6 &30 &10 &4.94 \\
\hline
\end{tabular}
\end{center}
\end{table}}}
\begin{figure*}
\centering{
\includegraphics[width=2\columnwidth]{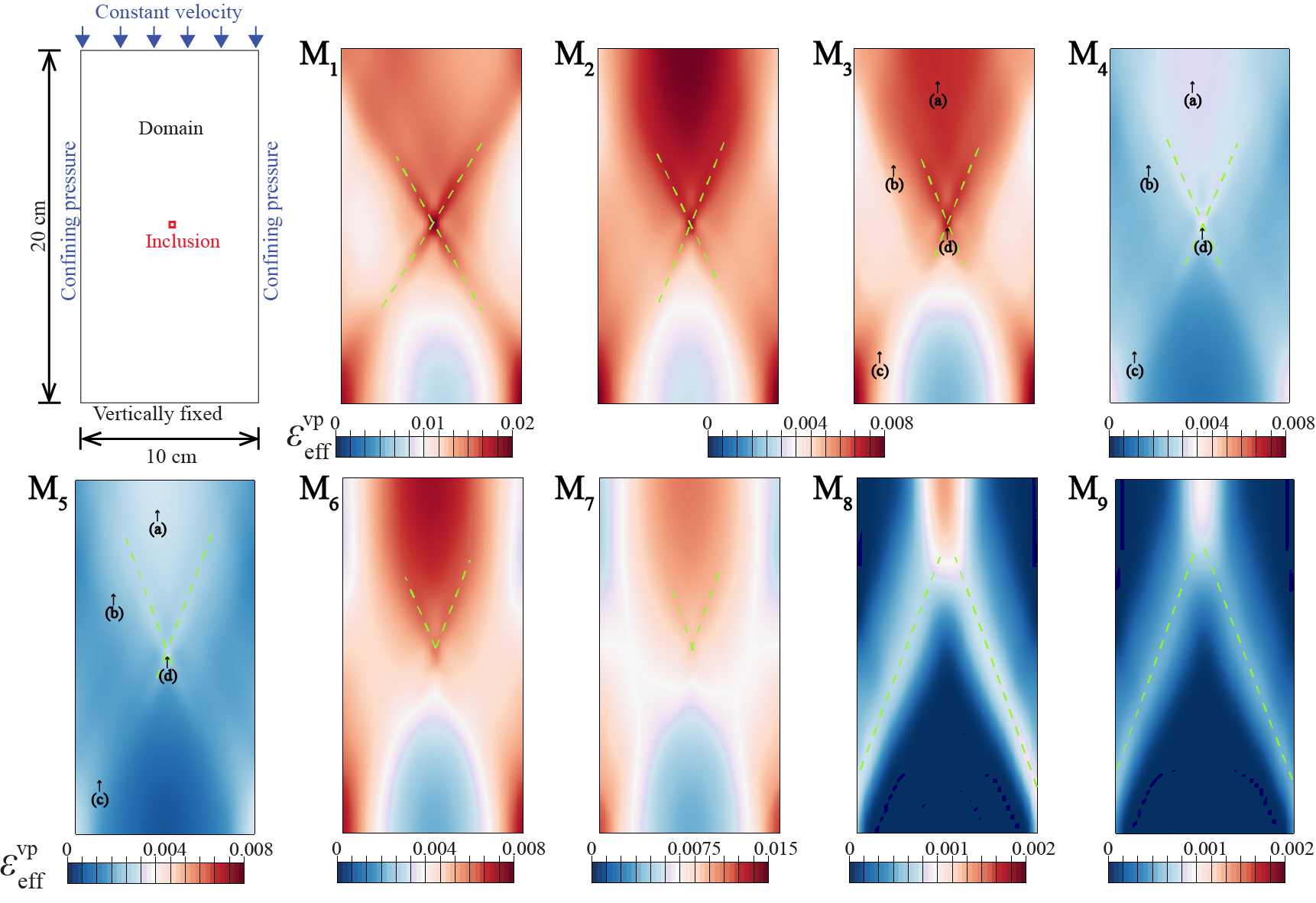}}
\caption{Parameter tests and verification of constitutive law using the configuration of Figure \ref{shearbandsetup}. All plots have been shown in undeformed configuration, i.e., the deformed shape is not shown for ease of comparing the nature of the shear bands. The angles of shear bands are measured along a straight line through the center of the band with respect to the horizontal. The white dashed lines are the paths through the observed shear bands along which the angles are measured. (a), (b) and (c) in M\textsubscript{3}, M\textsubscript{4} and M\textsubscript{5} are the markers from which the axial stress-strain histories are extracted for given integration points representing the part of the model that is less close to the top boundary, near the middle, at the site of the weak inclusion and near the bottom of the model. The extracted curves are shown in Figure \ref{stressstrain}.}
\label{leroyandortiz}
\end{figure*}
\begin{figure}
\begin{center}
\includegraphics[width=\columnwidth]{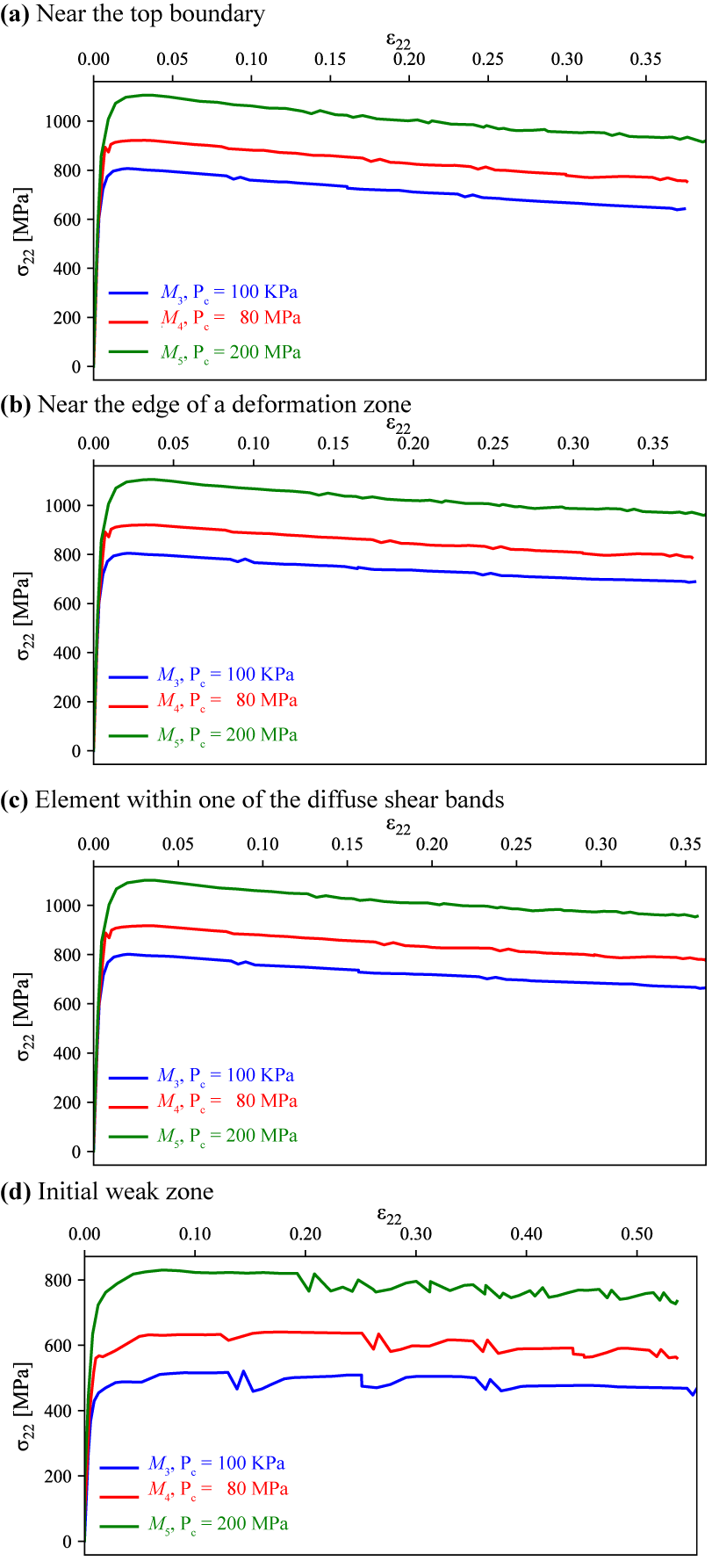}
\end{center}
\caption{Axial stress-strain curves from an element within the dominant shear bands and the initial weak element observed in models $M_3$, $M_4$ and $M_5$ for different confining pressures, respectively.}
\label{stressstrain}
\end{figure}
 
\begin{figure}
\begin{center}
\includegraphics[width=\columnwidth]{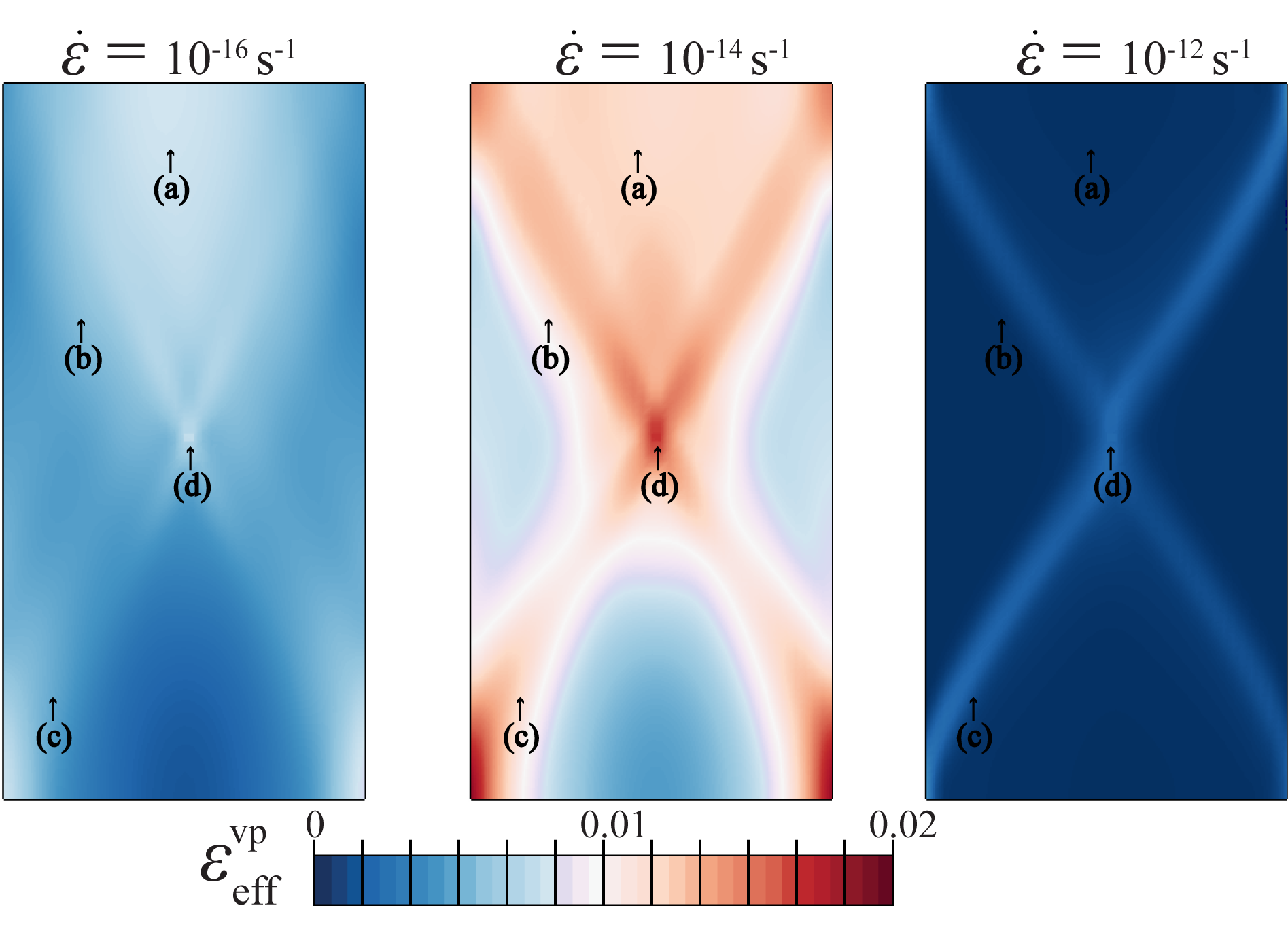}
\end{center}
\caption{Effect of applied strain rates on the deformation of the model M\textsubscript{5} at high temperature. a, b, c, d are markers for which axial stress-strain curves are extracted in Figure \ref{strainrate1D}.}
\label{strainrate2D}
\end{figure}
\begin{figure}
\begin{center}
\includegraphics[width=\columnwidth]{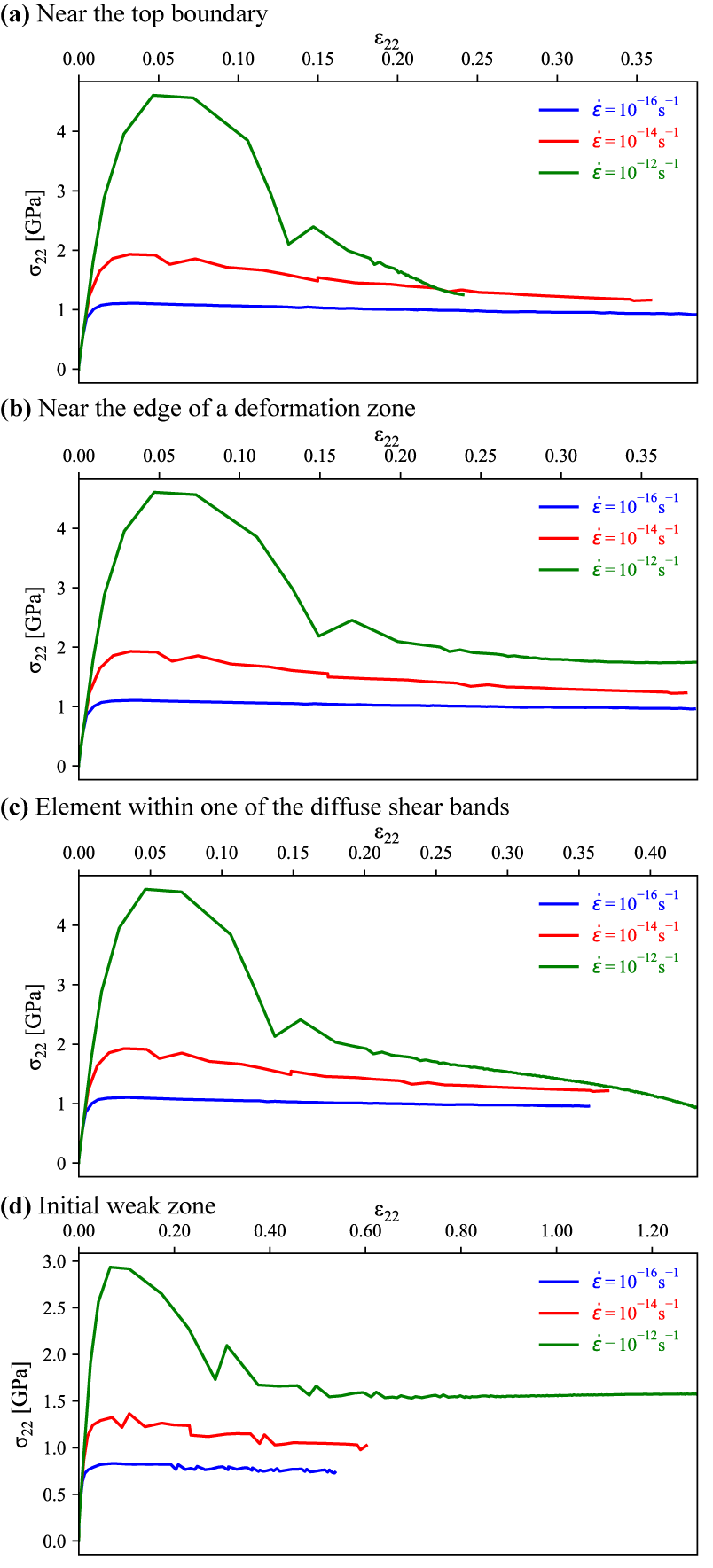}
\end{center}
\caption{Axial stress-strain curves to investigate the effect of different strain rates on the deformation.}
\label{strainrate1D}
\end{figure}
The effect of increasing confining pressure can be appreciated by comparing models M\textsubscript{3}, M\textsubscript{4} and M\textsubscript{5} in Figure \ref{leroyandortiz}. As previously highlighted, shear bands initialized as \enquote{X-shaped} bands subsequently developing into a dominant band efficiently localizing the deformation. The observation in Figure \ref{leroyandortiz} is that the intensity of deformation decreases with increasing confining pressure, and the rate of growth diminishes. This point is elucidated  as the elastic limit is higher with increasing confining pressure for four different integration points in elements close to the top boundary, near the edge of one of the most deformed zones, near the bottom of the model and the weak inclusion (Figure \ref{stressstrain}). In the four cases, the axial stress-strain history indicates that higher confining pressures lead to a a higher yield stress. Although our constitutive laws prescribe mechanical hardening as deformation increases, i.e., friction and cohesion hardening, the stress-strain curves for the different integration points exhibit progressive strain-softening for all confining pressures, indicating that localization is proceeding, ie. shear bands are becoming globally weaker as deformation proceeds. 

By comparing the results of model M\textsubscript{5} and M\textsubscript{6}, we can demonstrate the effect of increasing dilatancy. Although the deformation styles are similar, the amplitudes of deformation of model M\textsubscript{6} are clearly higher. Model M\textsubscript{6} can be compared to model M\textsubscript{7} showing the effect of a higher dilatancy in the weak element when compared with the rest of the model. Higher dilatancy thus seems to contribute a positive feedback which tends to enhance shear deformation. This shall be explored at a later section.

Model M\textsubscript{8} demonstrates a different pattern of deformation from the previous models with focused deformation along two shear bands whose disposition is such that their intersection develops at the top boundary rather than symmetrically about the weak element. Higher frictional and dilatancy parameters have been used for this test, which are values typically used for sand and geological materials. Model M\textsubscript{9} utilizes the same parameters as Model M\textsubscript{8} with a lower Young's modulus for the weak zone. The growth of the shear band is observed to be slower, compared to Model M\textsubscript{8}.

For the different material properties and loading conditions, the shear band angles ranged between 58${^\circ}$ and 72${^\circ}$. Differences are obvious in the magnitude,  growth-rate and thickness of the shear bands. The growth rate of the shear band seems to be slowed by high stress exponents and high confining pressures. Shear bands are enhanced by low confining pressures, low friction angle and high dilatancy angles; however, the combination of parameters, the range  between the parameters may be more important than the value of any given individual parameter. One can note generally several different kinds of top-vs-bottom asymmetry in the internal deformation of the sample domains.

Finally, we investigated the effect of applied strain rate to further test the robustness of our constitutive law. Shown in Figure \ref{strainrate2D} while Figure \ref{strainrate1D} shows the axial stress-strain curves. The figures indicate that the deformation indeed depends on the strain rates, with the stresses becoming notably higher in the tests with highest strain rates.
\begin{figure*}
    \centering
\includegraphics[width=\textwidth]{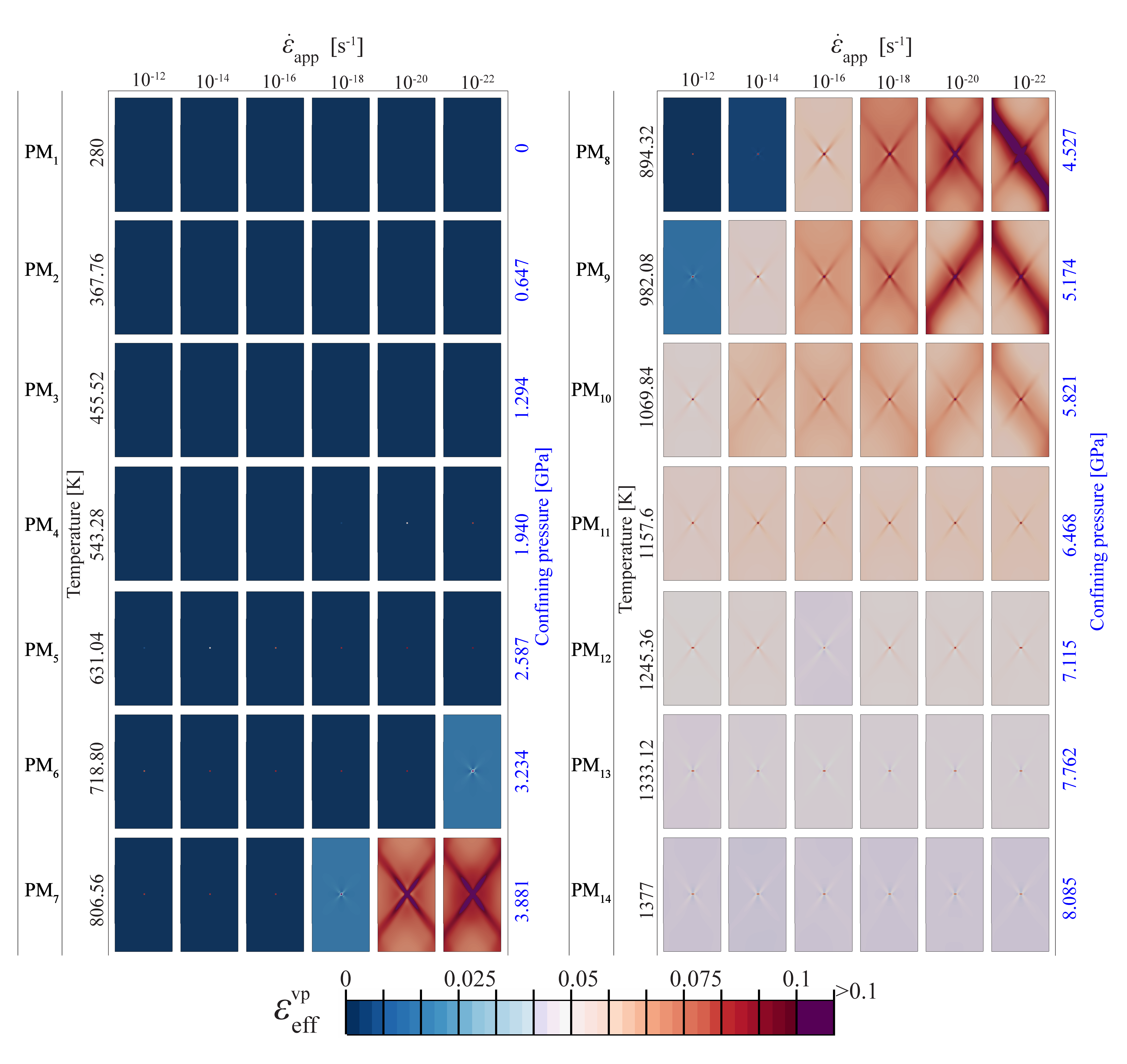}
    \caption{Results of biaxial compression tests using our custom constitutive law for different temperatures, confining pressures and strain rates. Left panel shows 0 to 120 km, while right panel shows 140 to 250 km. Temperatures and lithostatic pressures are shown on the left and right for the corresponding depths. Visco-plastic work is converted to heat.}
\label{thermodef}
\end{figure*}
\begin{figure}
\centering
\includegraphics[width=\columnwidth]{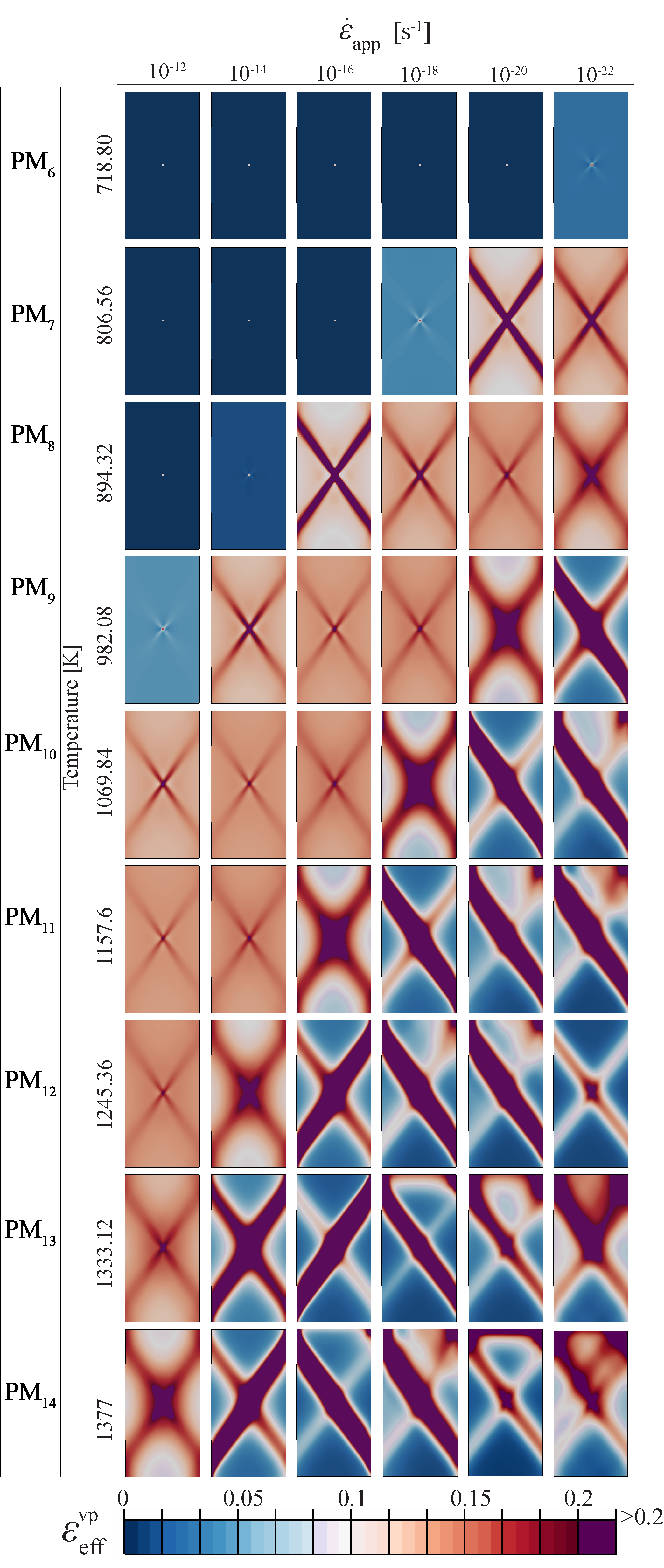}
\caption{Results of biaxial compression tests using our custom constitutive law for different temperatures, different applied strain rates and constant confining pressure. Visco-plastic work is converted to heat.}
\label{thermodef_lcp}
\end{figure}
\subsection{Temperature and pressure conditions for strain localization in elasto-thermo-viscoplastic deformation}
The above plane strain biaxial compression numerical experiments have been carried at constant temperature, a range of confining pressures appropriate for lithospheric deformation,  different mechanical properties and three different applied strain rates. In reality, temperature is hardly constant, a given rock unit may undergo different style of deformation depending on the surrounding conditions. These are important parameters and mechanisms to consider in compression-driven experiments \cite{leng2011,leng2015,kiss2020,auzemery2020,zhang2021,zhou2021}. 

Various attempts at investigating deformation in compression-driven experiments treat the rheology as visco-elastic (creeping at any stress), visco-plastic (creep is bounded by a plastic yield criterion), incompressible (neglecting volumetric plastic strains), e.g., \cite{thielmann2012}. The shear bands are not exactly the same depending on whether they developed in the ductile or the brittle regime. For example, we have seen from the benchmark tests in plasticity that the shear bands localize efficiently in brittle regimes and are found to be more diffuse in the ductile regime where temperature effects are a strong influence. Attempts to study the influences on shear bands formation in brittle regimes showed that the relationship between a prescribed heterogeneity and element size determines the shear band angles more than rheological properties and element types\cite{kaus2010}; but while these studies did not account for volumetric plastic strains, they also did not consider the influence of temperature which plays a role in the rheological description, hence the shear band formation. In the constitutive laws we described earlier, we adopt an elasto-thermo-viscoplastic rheology useful for frictional materials. Using this rheology, we now investigate the strain rates, temperature and pressure conditions that favour the localization of deformation. Towards this, we utilize some of the material properties already used for geodynamic subduction modelling in the literature. We present the results of compression of samples with similar loading conditions as would be expected in such geodynamic models, i.e., confining pressures (approximating a lithostatic stress state) and different convergence rates (applied strain rates) based on the expected range of geodynamic strain rates. The input model is the same with the homogeneous sample and a weak zone in the middle as shown in Figure \ref{shearbandsetup}.

In order to explore (at sample scale) the mechanisms of shear band formation under lithospheric conditions, we estimated confining pressures for gravity loading and corresponding temperatures are estimated using a thermal gradient of 4.39 K/km. The results of the compression for different applied strain rates are shown in Figure \ref{thermodef}. Based on the constitutive law, it is apparent that the contribution of viscoplastic flow to the irreversible deformation is infinitesimal at low temperatures and becomes observable at temperatures ${>700}$ K. Above this initial temperature, localization is most efficient at the lowest strain rates, but for initial temperatures above 1070K, localization is still observed even at the higher strain rates.

{\tiny{\begin{table*}
\begin{minipage}{158mm}
\caption{Material properties used for the model and the embedded weak nucleation element for the biaxial compression tests. Model: E - Young's modulus = 185 GPa, ${\nu}$ - Poisson ratio = 0.25, ${\rho}$ - density = 3300 kg.m${^{-3}}$, ${c_0}$ - initial cohesion = 20 MPa, ${A}$ - dislocation creep pre-exponential multiplier = 2.5${\times}$10${^{-17}}$ Pa${^{-m}}$s${^{-1}}$ (where ${m}$ is the stress exponent), ${E_a}$ - activation energy = 540 kJ.mol${^{-1}}$, and ${H}$ - hardening modulus = 50 MPa; weak element: E = 100 GPa, ${\nu}$ = 0.25, ${\rho}$ = 3300 kg.m${^{-3}}$, ${c_0}$ = 0.2 MPa, ${A}$ = 2${\times}$10${^{-21}} $Pa${^{-m}}$s${^{-1}}$, ${E_a}$ = 470 kJ.mol${^{-1}}$, and ${H}$ = 0.05 MPa}
\label{params}
\small
\begin{tabular}{@{}lccccccccccccr}
\hline
Model & Depth  & P${_c}$ & T & ${\varphi_i}$ & ${\varphi_f}$ & ${\psi_i}$ & ${m}$ & &  & ${\dot\varepsilon}_{app}$ & & & \\
 &  [km] &  [MPa]&  [K] & $[{^\circ}]$ & $[{^\circ}]$ & $[{^\circ}]$ &  & &  & [s\textsuperscript{-1}] &  & & \\
& &  &  & &   & &  &${10^{-12}}$ &${10^{-14}}$ &${10^{-16}}$ &${10^{-18}}$&${10^{-20}}$&${10^{-22}}$\\
\hline
PM\textsubscript{6} & &  &  & & &  & & & & & & &\\
\hspace{2mm} Model & 100 &  3234.00  &  718.80  &21 &37 &10 & 3.5 & \multirow{2}{*}{${0}$}&\multirow{2}{*}{${0}$}&\multirow{2}{*}{${0}$}&\multirow{2}{*}{${0}$}&\multirow{2}{*}{${0}$}&\multirow{2}{*}{${48.1^\circ}$}\\
\hspace{2mm} Element & 100 &  3234.00  &  718.80 &15 &15 &10 &4 & & & & &  & \\
\hline
PM\textsubscript{7} & &  &  & & &  & & &  & & & &\\
\hspace{2mm} Model & 120 &  3880.80  &  806.56  & 21& 37&10 & 3.5 & \multirow{2}{*}{${0}$}&\multirow{2}{*}{${0}$}&\multirow{2}{*}{${0}$}&\multirow{2}{*}{${46.4^\circ}$}&\multirow{2}{*}{${56.3^\circ}$}&\multirow{2}{*}{${54.8^\circ}$}\\
\hspace{2mm} Element & 120 &  3880.80  &  806.56  & 15&15 &10 & 4 & & & & & & \\
\hline
PM\textsubscript{8} & &  &  & & &  & & & & & & &\\
\hspace{2mm} Model & 140 &  4527.60  &  894.32  &21& 37&10 & 3.5 & \multirow{2}{*}{${0}$}&\multirow{2}{*}{${0}$}&\multirow{2}{*}{${51^\circ}$}&\multirow{2}{*}{${53.7^\circ}$}&\multirow{2}{*}{${55.1^\circ}$}&\multirow{2}{*}{${54^\circ}$}\\
\hspace{2mm} Element & 140 &  4527.60  &  894.32  & 15&15 &10 &4& & & & & &\\
\hline
PM\textsubscript{9} & &  &  & & &  & & & & & & &\\
\hspace{2mm} Model & 160 &  5174.40  &  982.08  & 21&37 &10 &3.5& \multirow{2}{*}{${43.8^\circ}$}&\multirow{2}{*}{${49.6^\circ}$}&\multirow{2}{*}{${54.1^\circ}$}&\multirow{2}{*}{${55.3^\circ}$}&\multirow{2}{*}{${55.1^\circ}$}&\multirow{2}{*}{${54.2^\circ}$} \\
\hspace{2mm} Element & 160 &  5174.40  &  982.08 &15 & 15& 10&4  & & & & & & \\
\hline
PM\textsubscript{10} & &  &  & & & & & &  & & & &\\
\hspace{2mm} Model & 180 &  5821.20  &  1069.84 &21 &37 &10 &3.5 &\multirow{2}{*}{${49.4^\circ}$}&\multirow{2}{*}{${52.6^\circ}$}&\multirow{2}{*}{${53^\circ}$}&\multirow{2}{*}{${52.3^\circ}$}&\multirow{2}{*}{${51.8^\circ}$}&\multirow{2}{*}{${54.8^\circ}$}\\
\hspace{2mm} Element & 180 &  5821.20  &  1069.84 &15 &15 &10 &4 & & & & & & \\
\hline
PM\textsubscript{11} & &  &  & & &  & & & & & & &\\
\hspace{2mm} Model & 200 &  6468.00  &  1157.60 &21 &37 &10&3.5 & \multirow{2}{*}{${53^\circ}$}&\multirow{2}{*}{${52.1^\circ}$}&\multirow{2}{*}{${48.4^\circ}$}&\multirow{2}{*}{${53.6^\circ}$}&\multirow{2}{*}{${52.6^\circ}$}&\multirow{2}{*}{${51.9^\circ}$}\\
\hspace{2mm} Element & 200 &  6468.00  &  1157.60 &15 &15 &10 &4& & & & & & \\
\hline
PM\textsubscript{12} & &  &  & & & & & &  & & & &\\
\hspace{2mm} Model & 220 &  7114.80  &  1245.36 & 21&37 &10&3.5 & \multirow{2}{*}{${54.7^\circ}$}&\multirow{2}{*}{${52.5^\circ}$}&\multirow{2}{*}{${52.2^\circ}$}&\multirow{2}{*}{${53.6^\circ}$}&\multirow{2}{*}{${52.2^\circ}$}&\multirow{2}{*}{${52.4^\circ}$}\\
\hspace{2mm} Element & 220 &  7114.80  &  1245.36 &15&15 &10 & 4 & & & & & & \\
\hline
PM\textsubscript{13} & &  &  & & &  & & & & & & &\\
\hspace{2mm} Model & 240 &  7761.60  &  1333.12 &21 &37 &10&3.5 & \multirow{2}{*}{${56^\circ}$}&\multirow{2}{*}{${53.3^\circ}$}&\multirow{2}{*}{${53.6^\circ}$}&\multirow{2}{*}{${53.7^\circ}$}&\multirow{2}{*}{${55.3^\circ}$}&\multirow{2}{*}{${52.1^\circ}$} \\
\hspace{2mm} Element & 240 &  7761.60  &  1333.12 &15 &15 & 10&4& & &  & & & \\
\hline
PM\textsubscript{14} & &  &  & & & & & &  & & & &\\
\hspace{2mm} Model & 250 &  8085.00  &  1377.00 &21&37 & 10 &3.5 & \multirow{2}{*}{${55.5^\circ}$}&\multirow{2}{*}{${54.4^\circ}$}&\multirow{2}{*}{${52.2^\circ}$}&\multirow{2}{*}{${56^\circ}$}&\multirow{2}{*}{${51.6^\circ}$}&\multirow{2}{*}{${52.5^\circ}$} \\
\hspace{2mm} Element & 250 &  8085.00  &  1377.00 &15 &15 &10 &4 & &  & & & \\
\hline
\end{tabular}
\end{minipage}
\end{table*}}}

Although shear bands of various intensities developed at high temperatures,  the observation already made (Figure \ref{stressstrain}) that increasing confining pressure attenuates shear band formation (due to the higher elastic limit) remains true; hence in Figure \ref{thermodef}, the bands are less prominent at the highest confining pressures explored. This implies that the initial geothermal gradient will play an important role. 

Considering that pressure and temperature increase simultaneously and increasing temperature is expected to enhance creeping effects which is apparently countered by increasing confining pressure, we repeat the experiment but with a constant and low confining pressure. The results are shown in Figure \ref{thermodef_lcp} where shear bands can be observed at high temperatures and (${>}$700 K) with efficient deformation increasing at higher strain rates for higher temperatures. The combined results shown in Figures \ref{thermodef} and \ref{thermodef_lcp} seem to imply that (in the natural system we are envisaging), the following evolution might be anticipated, i.e., the lithosphere should begin to heat up first where the shear bands nucleate, presulably near its base where temperature is highest. But as isotherms move upwards towards the surface, the deformation bands should become progressively more pronounced aiding the development of lithospheric-scale shear zones, which are required in order to initiate the subduction process.

\subsection{Shear heating versus dilatational contributions to heating}\label{shearvsvol}
Conventional deformational heating utilizes the deviatoric stress terms and viscoplastic strain rates to compute the heat source term in a so-called \enquote{shear heating} \cite{babeyko2008,duretz11,willis2019}. The formulation of our constitutive law accounts for dilatational deformation. Therefore,  we were able to carry out simulations in which our conservation of energy equation incorporates a heat source terms accounting for both dilatational and shear deformations or leaving out the former. In these simulations, we have used a constant initial temperature of 1318 K chosen due to the presence of shear bands at all applied strain rates  as can be inferred from Figure \ref{thermodef} to study the contribution of shear heating alone and combined shear/dilatational heating.

\begin{figure}
    \centering
\includegraphics[width=4.5cm]{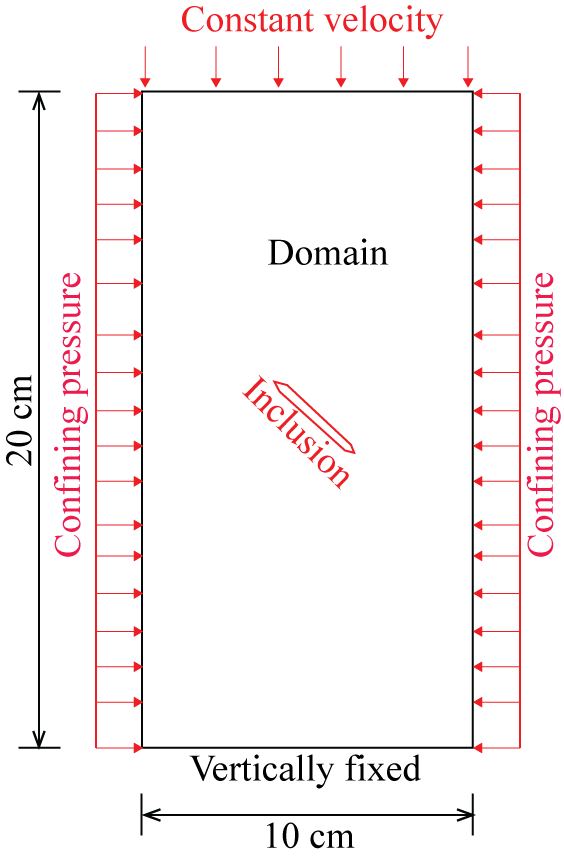}
\caption{Model set up with weak inclusion inclined at an angle and encompassing a few elements. }
\label{newsetup}
\end{figure}

In order to compute just shear heating, we utilize only the deviatoric stress components and the viscoplastic strain rates to compute the heat source term, whereas when computing both shear and dilatational heating, we utilize the total stress tensor with the viscoplastic strain rate components to compute the heat source term. The input model is shown in Figure \ref{newsetup}, where the weak zone has been inclined at an angle and incorporates more that an element. Even though we have seen that a single element can in fact lead to the development of shear bands; for the tests forthwith, we typically use weak zones with multiple elements, as is usually done in large scale problems is adopted.   

\begin{figure*}
    \centering
\includegraphics[width=1.5\columnwidth]{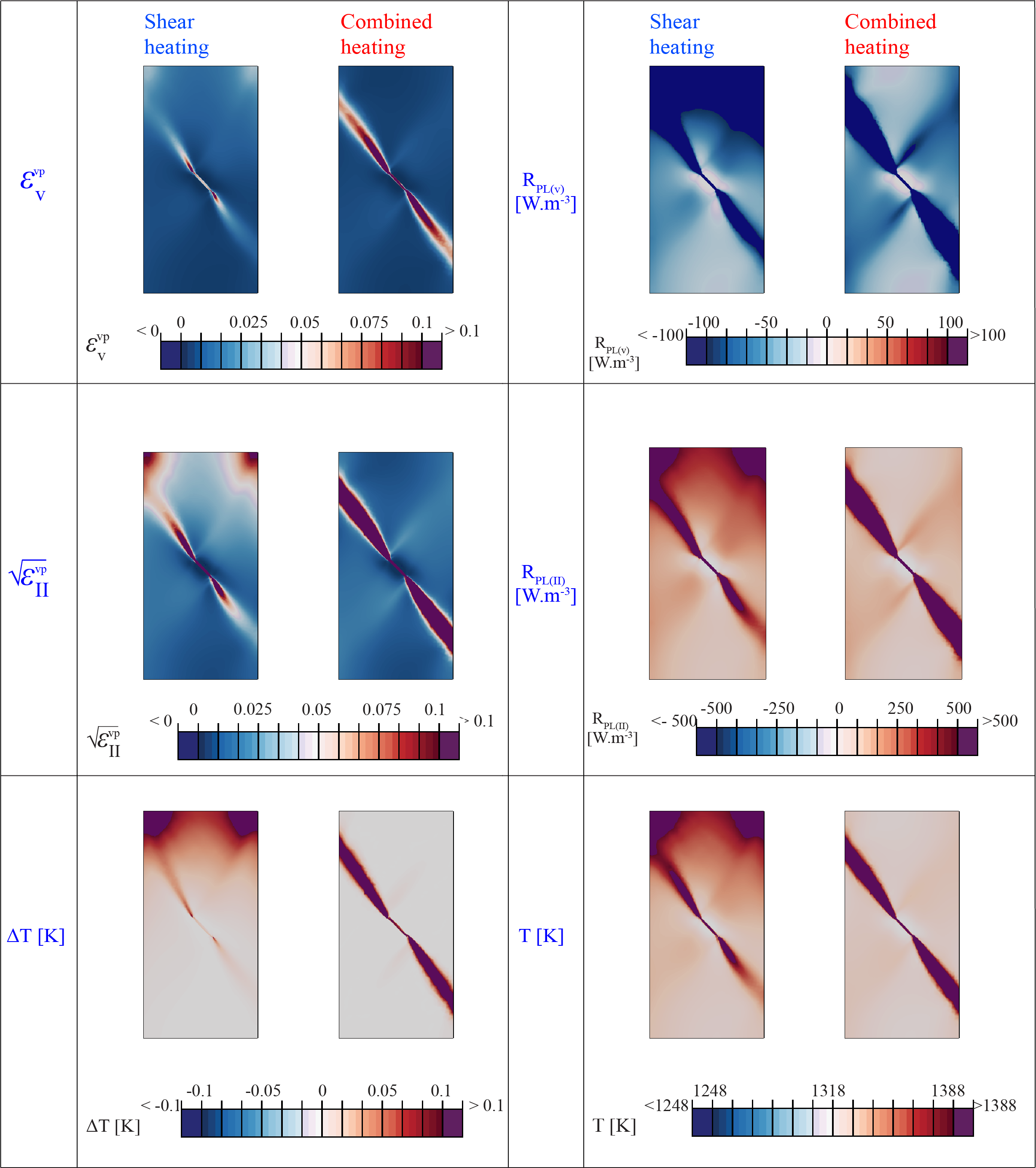}
    \caption{The comparison of shear deformation and combined shear with volumetric deformation modes showing the volumetric viscoplastic strain invariant (${\varepsilon_{\textrm{v}}^\textrm{vp}}$), deviatoric viscoplastic strain invariant (${\varepsilon_\textrm{II}^\textrm{vp}}$), associated volumetric- (${\textrm{R}_\textrm{{PL(v)}}}$) and shear heat generation per unit volume (${\textrm{R}_\textrm{{PL(II)}}}$), instantaneous temperature change (${\Delta{\textrm{T}}}$) and integrated temperature ($\textrm{T}$). }
\label{heatgen}
\end{figure*}

We examine volumetric viscoplastic strain invariants, deviatoric viscoplastic strain invariants, instantanteneous temperature change, volumetric and shear heat generation; and accummulated temperature in Figure \ref{heatgen}. The influence of dilatancy on the deformation is apparent: localization zones are narrower when shear components are utilized as the sole contributors to the heat source, but wider and continuous when the full stress tensor is used, i.e., incorporating dilatancy effects. Another observation concerns deformation at the boundaries, which is predominant when shear contributions are utilized for heating. Futhermore, although dilatational heat generation is negative, the deviatoric stress contributions to heating are observed to be up to five times higher and positive. We also infer from Figure \ref{heatgen} that development of localized shear deformation is somehow enhanced by the inclusion of dilatancy, and the net effect is to enhance and slightly broaden the heated zone when the full stress tensor is used. Deformation focuses on somewhat narrower shear bands and at the boundary when shear components are used. 

\begin{figure*}
    \centering
\includegraphics[width=1.5\columnwidth]{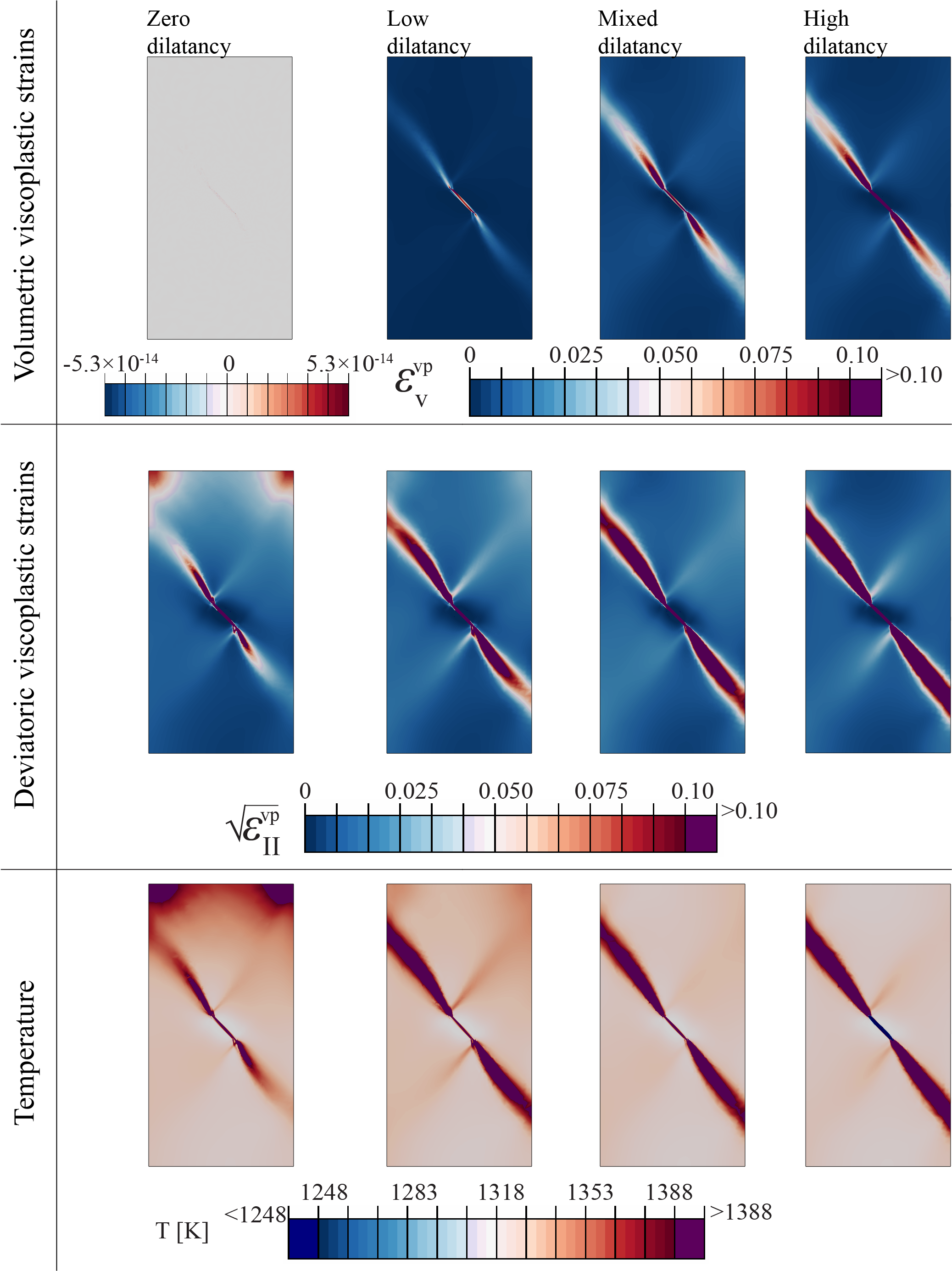}
    \caption{The role of dilatancy in deformational and heating. High dilatancy in our numerical experiment is ${\psi}=10^\circ$ for the model and the weak zone; low dilatancy is ${\psi} = 2^\circ$; mixed dilatancy assigns the ${\psi}=10^\circ$ for the model and ${\psi}=2^\circ$ for the weak zone; and zero dilatancy treats the deformation as volume-preserving deformation.}
\label{dilate}
\end{figure*}

Dilatancy has been reported as the critical parameter describing increase in volume (dilation) in a material undergoing shear distortion, and for the case of simple shear, it is estimated as the ratio of the plastic volumetric strain rate and the plastic shear strain rate \cite{vermeer1984,vermeer1990,hobbs1990,poliakov1993,alonso05}. The parameter used to account for how much a material dilates is the dilatancy angle. \cite{hobbs1989} argue that the magnitudes of friction and dilation angles influence the width of shear bands, which appears consistent with our results in Figure \ref{heatgen}. The influence of this parameter has not been explored in detail, and models that attempt to utilize the parameter treat it either by assuming it is zero (theoretically, volume preserving deformation) or associative \cite{alonso05}, which is not a typical property of rocks. 

To explore the quantitative influence of different dilatancy angles in deformation studies, we ran simulations with zero-dilatancy angles (visco-plastically volume-preserving deformation), low dilatancy angle, high dilatancy angle and mixed dilatancy angle. As shown in Figure \ref{dilate}, zero-dilatancy leads to a negligible volumetric viscoplastic strain (of the order of 10\textsuperscript{-14}) while increasing dilatancy leads to corresponding increasing volumetric viscoplastic strains. The same behaviour of increasing viscoplastic deviatoric strain magnitude with increasing dilatancy is observed, but there is some deviatoric deformation at zero dilatancy which appear as narrow focused shear-bands attenuating towards the boundaries of the model, with some deformation from the top corners. High dilatancy angles lead to a wider deformation zone with less amplitude in the deformation zone. The thermal feedback follows the same pattern as the viscoplastic deviatoric strain invariant. As we already saw in Figure \ref{heatgen}, the magnitude of the heat generated from viscoplastic deviatoric deformation can be up to five times more than the volumetric heat generation rate; it is therefore not surprising that the accumulated temperature follows the deviatoric viscoplastic deformation. 

\subsection{Influence of the initial orientation of seed zone on shear bands}
Various authors have shown that a combination of friction and dilatancy angles theoretically contribute to  influencing the angles for which shear bands are mechanically stable. For example, the Coulomb angle theoretically depends on the friction angle, while the Roscoe angle depends on the dilatancy angle and the Arthur angle depends on the combination of the friction and dilatancy angles \cite{coulomb1773,roscoe1970,arthur1977}. 

Numerical studies also show that shear bands in plasticity are controlled by a ratio between the size of an initial seed (nucleation) zone and the size of the numerical mesh chosen \cite{kaus2010}, indicating that the shear bands are dependent on the geometry of the nucleation zone and mesh resolution. \cite{rudnicki1975} investigated the idea that localization (shear bands) may depend on an instability in the choice of constitutive description (and plasticity parameters like the friction and dilatancy angles, and the hardening modulus). \cite{rice1976} also reviewed conditions favouring shear band localization with emphasis on non-associative flow, non-uniformity in material properties and vertex-like yielding. \cite{poliakov1993} showed the influence of the frictional and dilatational properties of overburdens on the formation and propagation of shear bands both in extensional and compressional regimes, assuming the domain is underlain by an inviscid fluid.  Other weakening mechanisms like grain-size reduction \cite{thielmann2015} and lubrication by water \cite{lieb2001} have been investigated to understand their role in providing positive feedbacks for rapid conversion of deformational work to heat and initiating subduction, respectively. 

One of the influential geometrical parameters, during the initiation and development of subduction, is expected to be a plastically-weak shear zone; and as a prelude towards the anticipated subduction thrust, a few authors orient the initial seed zone thereto, e.g., \cite{ruh2015}. We take note that imposing a convergence on one of the boundaries leads to a steepening of the seed zone and may not fit with the expected subduction angle. However, this is commonly compensated for by the choice of boundary conditions, where for example, there is a convergent inflow at the top half of the model and an outflow at the bottom half leading to a zero net output \cite{ruh2015}. Here, we investigate (still on the small-scale model) the influence of the orientation of the initial seed zone on the orientation of the shear bands (which are sometimes interpreted as faults in lithospheric contexts) that eventually develop.

\begin{figure}
    \centering
\includegraphics[width=8cm]{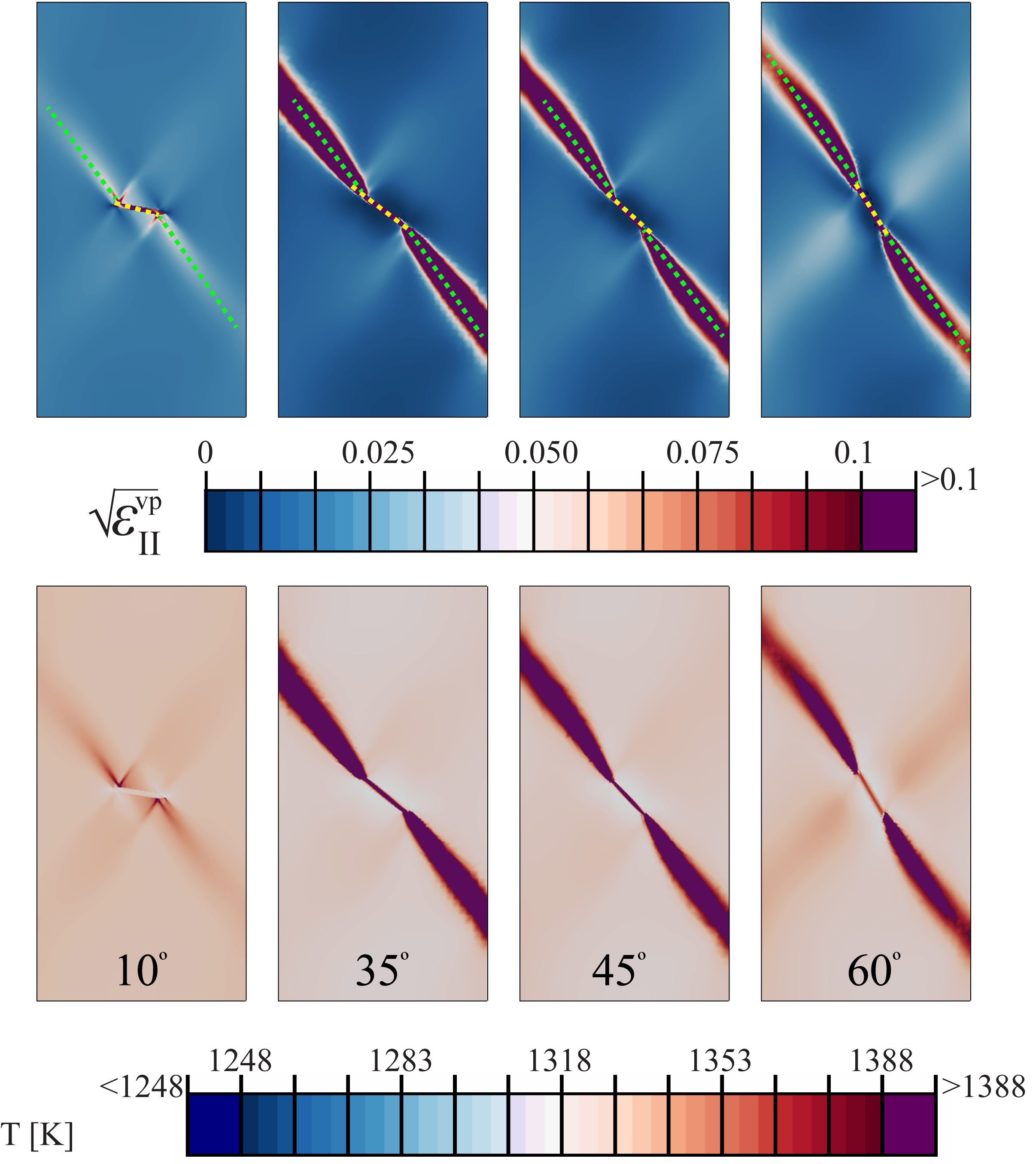}
    \caption{The role of orientation of initial seed zone. The applied strain rate is 5${\times}10^{-12}\;s^{-1}$ and a temperature of 1318 K chosen due to the unambiguous observation of shear bands. The yellow dashed lines follow the core of the heterogeneity while the green dashed lines highlight the core of the shear band.}
\label{angleinfluence}
\end{figure}
We consider four different orientations of the initial weak zone: 10${^\circ}$, 35${^\circ}$, 45${^\circ}$ and 60${^\circ}$ oriented North-West South-East at the center of the model (Figure \ref{newsetup}). Our observations are shown in Figure \ref{angleinfluence}, and they indicate that irrespective of the orientation of the initial seed zone, the angles of shear bands range between ${\sim}$53 - 55${^\circ}$. Localization seems more efficient for initial orientations of 35${^\circ}$ and 45${^\circ}$, and the thermal localization follows the same pattern. However, the two cases of 10${^\circ}$ do not show sufficient evidence for efficient localization compared to those with higher initial orientations of the weak inclusion. The thermal feedback is also not as efficient. In the case of 60${^\circ}$ initial orientation of the seed zone, the shear bands seem to attenuate as they approach the boundaries, and the same effect is observed in the thermal evolution (Figure \ref{angleinfluence}).

\subsection{The role of thermal diffusivity}

The experiments presented above have been done at zero thermal diffusivities. Real materials have finite diffusivities whose values will contribute to the development of a thermal steady state, because heat produced will tend to be evacuated by diffusion. Diffusivity will clearly have some feedback effect on rheology via the temperature field. In the next experiments, we examine the role of diffusivities in the deformation. We utilize the input shown in Figure \ref{newsetup}. 
\begin{figure*}
    \centering
\includegraphics[width=14cm]{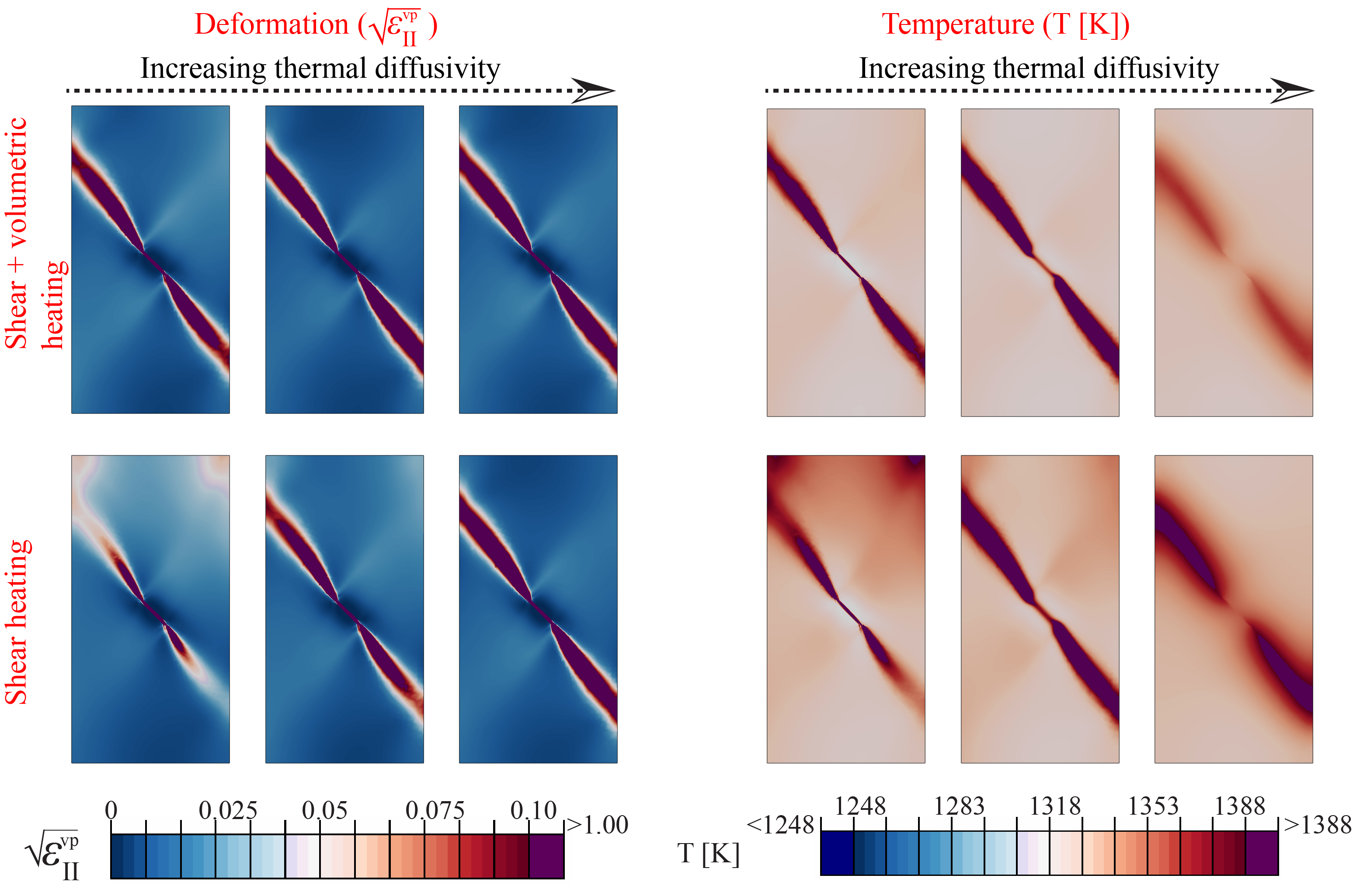}
    \caption{The competition between volumetric heating/cooling and shear heating and the role of increasing thermal conductivity. Applied strain rate is 5${\times}10^{-12}\;s^{-1}$ and a constant initial temperature of 1318 K chosen due to the unambiguous observation of shear bands. The confining pressure is 2.587 GPa.}
\label{thermdiff}
\end{figure*}

For zero thermal diffusivity, the most deformation is seen for the combined volumetric and shear deformation as shown in Figure \ref{thermdiff}, outlined by wider and continuous shear bands, while the narrower deformation zones are observed in the shear heating alone. Increasing thermal diffusivity exhibits similar deformation for both cases. However, the influence of the thermal diffusivities is even more apparent in the thermal evolution. Zero and low thermal diffusivities lead to focused shear bands while the highest diffusivities lead to a more diffuse heated zone, somewhat wider than the deformed area (Figure \ref{thermdiff}). 

\section{Application to a geodynamic problem: the case of a potential subduction initiation}
\begin{figure*}
    \centering
\includegraphics[width=13cm]{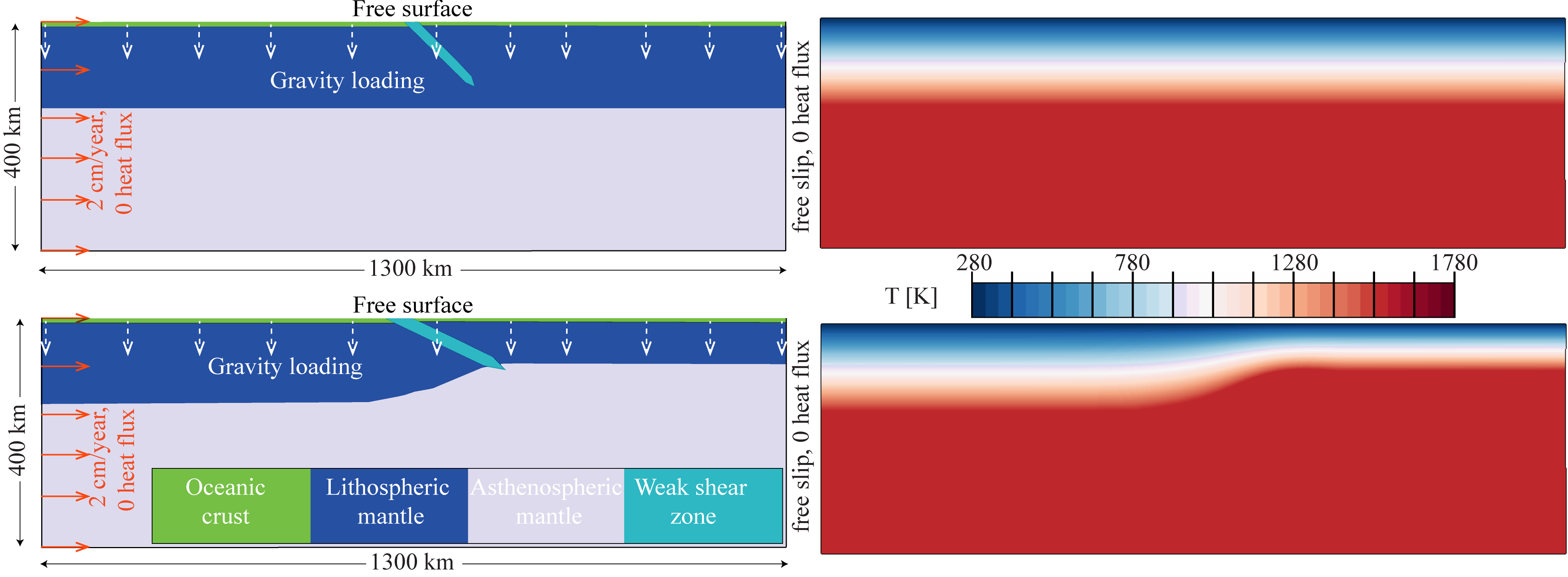}
    \caption{Input models with indicators for rock compositions and kinematic conditions. The right panel shows the corresponding initial temperature field, with a 1-D profile from the surface, and a thermally and mechanically thick impending subducting slab versus a thinner impending overriding plate.}
\label{input_geo}
\end{figure*}

To investigate the possible thermomechanical conditions that may lead to the initiation of a subduction zone using our constitutive law, we set up a geodynamic problem with initial conditions shown in Figure \ref{input_geo}. We investigate two scenarios: (1) a laterally uniform initial temperature and density with shear and volumetric heating with a 1-D temperature profile as input, (2) same as (1), but with a lateral variation in initial plate thickness. The material properties utilized are shown in Table \ref{tabb}. 
{{\begin{table*}
\begin{minipage}{160mm}
\caption{Rheological description of the lithologies in the geodynamic model}
\label{tabb}
\begin{tabular}{@{}llllllll}
\hline
Material property  (unit):  & Property name & Asthenospheric  & Lithospheric & Weak shear & Crust\\
&                 &               mantle  &  mantle         & zone \\
\hline
Mechanical &     &        &      &  \\[2pt]
\hspace{2mm}E (GPa)&Young's modulus     & 185       & 185     & 100 & 100\\[2pt]
\hspace{2mm}$v$   & Poisson's ratio  & 0.25       & 0.25     & 0.25 & 0.25 \\[2pt]
\hspace{2mm}$K {\;} (GPa)$ & Bulk modulus     & 123       & 123     & 67 & 67 \\[2pt]
\hspace{2mm}$G {\;} (GPa)$ & Shear modulus     & 74       & 74     & 40 & 40\\[2pt]
\hspace{2mm}$\rho (kg.m^{-3})$ & Density    & 3200       & 3300     & 3300 & 2900\\[2pt]
\hspace{2mm}$ g (m.s^{-2})$ & Gravitational constant   &9.8       & 9.8     & 9.8& 9.8
\\[2pt]
\hline
Thermal &     &        &      &  \\[2pt]
\hspace{2mm}$C_p{\;}(J.kg^{-1}.K^{-1})$ & Specific heat capacity & 1200       & 1200& 1200     & 1200\\[2pt]
\hspace{2mm}$T_{ref} {\;}(K)$ & Reference temperature & 280    & 280     & 280 & 280\\
\hspace{2mm}$\alpha_\textrm{th} {\;}(m^2s^{-1})$ & Thermal diffusivity & ${8.3\times{10^{-7}}}$  & ${8.1\times{10^{-7}}}$     & ${8.1\times{10^{-7}}}$ & ${9.2\times{10^{-7}}}$\\
\hline
Dislocation creep &     &        &      &  \\[2pt]
\hspace{2mm}$A \; (Pa^{-m}.s^{-1})$ & Pre-exponential multiplier    & $4.85\times10^{-17}$       & $4.85\times10^{-17}$   & $2\times10^{-21}$ & $2.08\times10^{-23}$\\[2pt]
\hspace{2mm}$E_a\; (kJ.mol^{-1})$ & Creep activation energy    & 470       & 540     & 470 & 238 \\[2pt]
\hspace{2mm}$R\; (J.K^{-1}.mol^{-1})$ & Molecular gas constant    & 8.31  & 8.31     & 8.31 & 8.31 \\
\hline
Plasticity &     &        &      &  \\[2pt]
\hspace{2mm}$\phi_i {\;} (^o)$& Initial friction angle     & 23.6       & 23.6     & 2.9 & 23.6\\
\hspace{2mm}$\phi_f {\;} (^o)$& Saturation friction angle     & 36.9       & 36.9     & 5.7 & 30\\
\hspace{2mm}$\psi {\;} (^o)$& Dilatancy angle     & 10       & 10     & 10 & 2\\[2pt]
\hspace{2mm}$c_0 {\;} (MPa)$& Initial cohesion     & 20       & 20     & 0.2 & 0 \\[2pt]
\hspace{2mm}$H {\;} (MPa)$ &  Hardening modulus    & 5       & 5     & 0 & 5 \\[2pt]
\hspace{2mm}$m $ & Stress exponent    & 3.5       & 3.5     & 4 & 3.2 \\[2pt]
\hline
\end{tabular}
\end{minipage}
\end{table*}}}

\begin{figure*}
    \centering
\includegraphics[width=13cm]{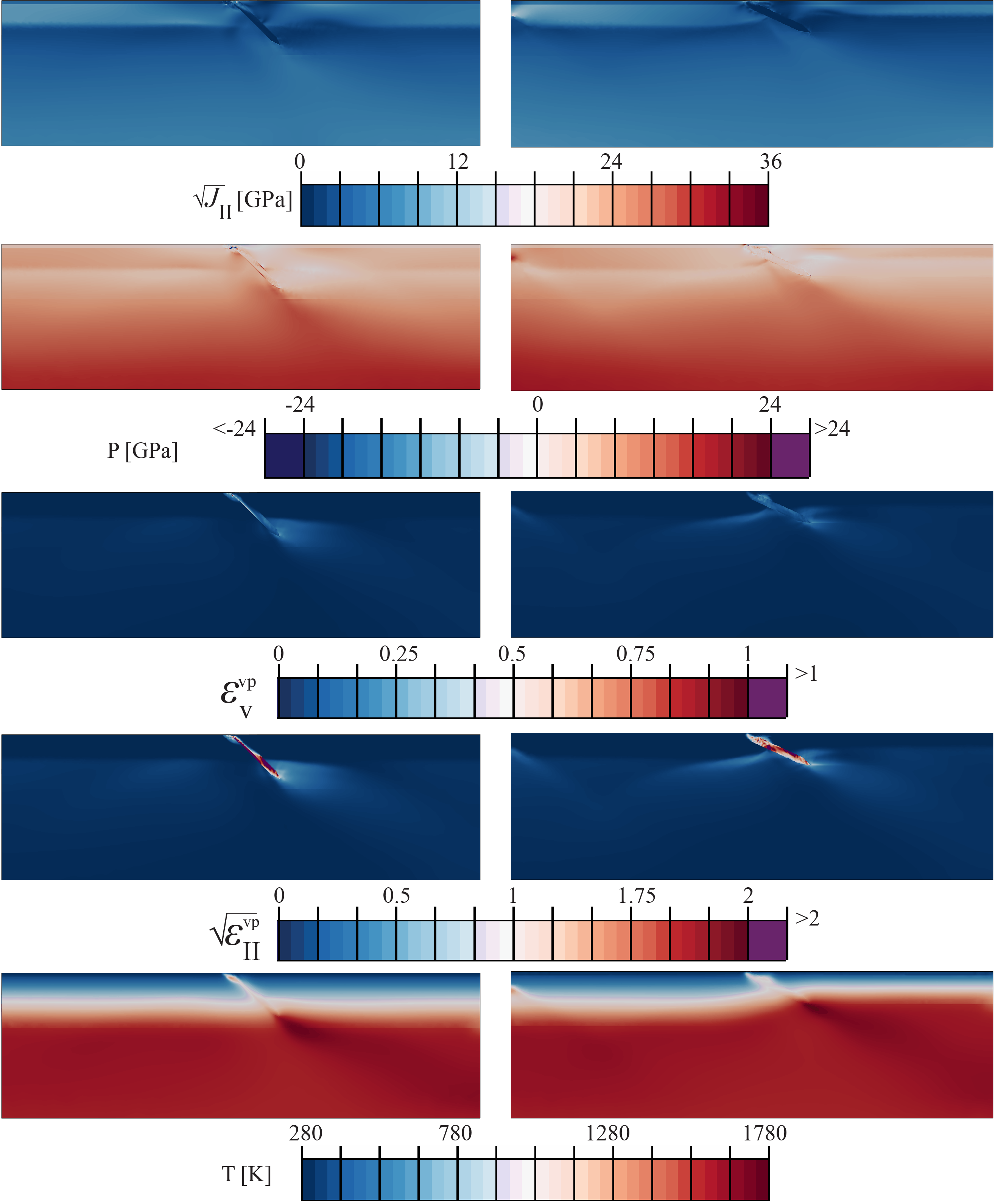}
    \caption{2-D model results showing the deviatoric stress invariant, pressure, volumetric viscoplastic strain invariant, deviatoric viscoplastic strain invariant and temperature. These results have been shown in undeformed configuration to compare deformation styles for the two scenarios.}
\label{input_geo_results}
\end{figure*}

\begin{figure}
    \centering
\includegraphics[width=8cm]{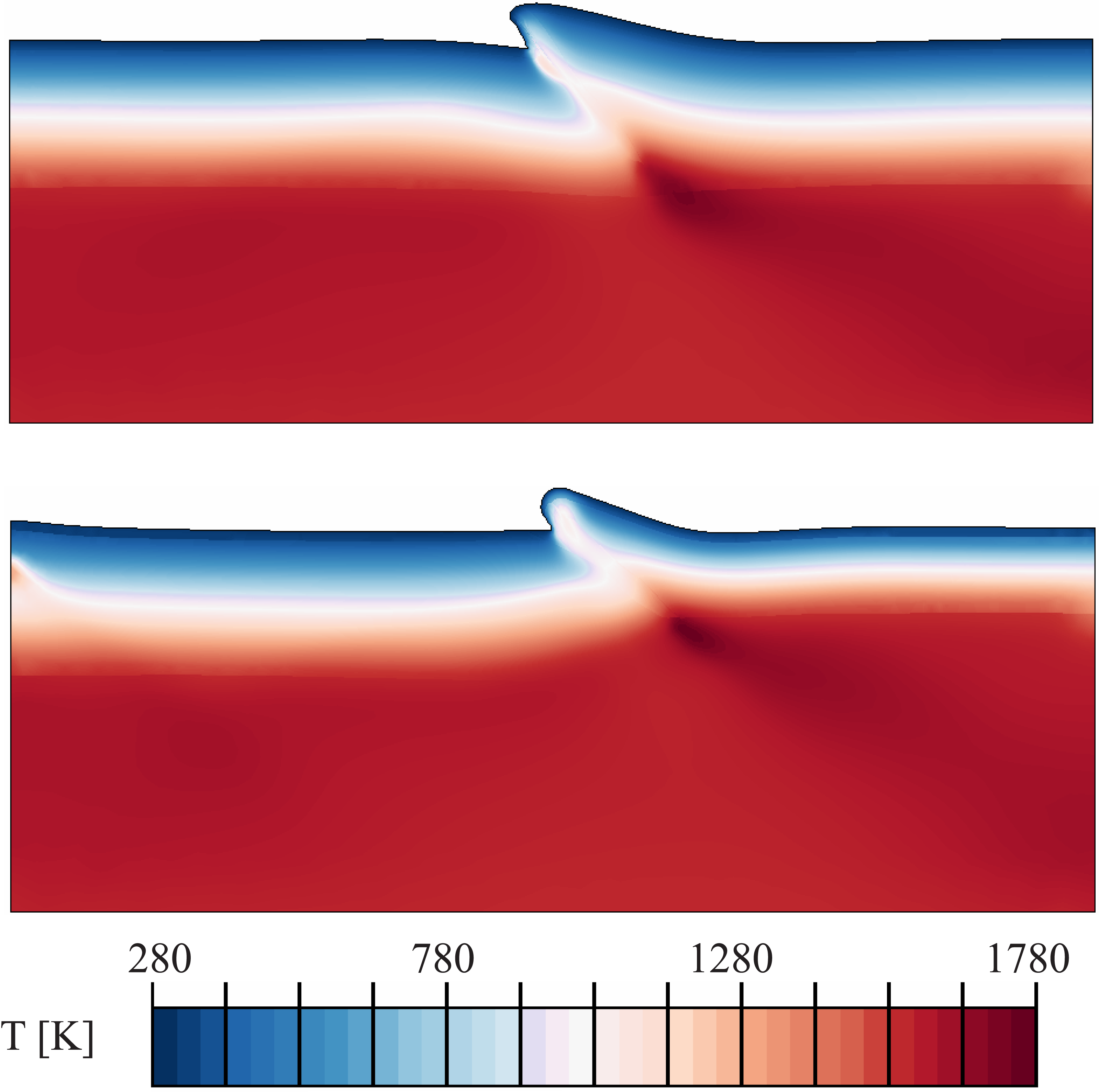}
    \caption{Temperature perturbations of two case studies of a uniform lateral temperature and laterally varying temperature inputs shown in deformed configuration.}
\label{deformed_config}
\end{figure}

The magnitudes of stresses (deviatoric stress invariant and pressure), volumetric viscoplastic strains, deviatoric viscoplastic strains, and temperature are shown in Figure \ref{input_geo_results}. The localization of deformation and growth of the localization zone is apparent at the sub-lithospheric tip of the weak zone. The zone of intense deformation is concentrated around the interface between the weak shear zone and the lithosphere, where the most shear deformation is observed. Although the temperature rise at this time interval is a modest ${\sim}$200 K, compared to the input model, the zones where the temperature increase is most apparent are just beneath and stretching forwards from the sub-lithospheric tip of the weak zone. There is also a hint of a cold \enquote{slab} with a downgoing trend of the temperature field in the case of the input with a laterally uniform temperature. This observation is shown in Figure \ref{deformed_config}. The overall shortening is ${\sim}$140 km, i.e., about 10${\%}$ after about ${\sim}$6.7 million years with observed strain rates lying between ${10^{-16}\;s^{-1}}$ and ${10^{-13}\;s^{-1}}$.

As we observed in the laboratory-scale numerical experiments, the creep contribution becomes apparent at temperatures  of 700K and above. Sunsequently, the development of the shear band from the initial seed is then influenced by diffusive processes, and by the feedback to rheology as deformation raises further the temperature. Although the topography generated in these simulations appears unrealistic, this may be explained by the absence of erosive processes in the constitutive description. We have also not imposed any viscosity, ab initio, in the material properties.
\section{Discussion}
\subsection{Shear band localization in viscoplasticity}
Among the most critical factors in compression-driven numerical models are temperature, loading rate and numerical grid size. In our calculations, we have attempted to resolve some of the influences of rheological parameters, loading conditions, thermal conditions and the interplay between mechanical deformation and thermal feedback to provide insight into the combination of properties and mechanisms favouring or opposing localization of deformation from a solid-mechanical viewpoint.

We have also introduced a novel feature into the computational scheme. Our approach has been to test our constitutive law on plasticity problems for frictional materials and extended to problems of strain localization on small samples by considering temperature effects, loading rates, material properties and geometrical properties.

Our simulations show that localization takes place in the form of shear band development at temperatures in excess of approximately 700 K. Moreover, localization proceeds more efficiently when the confining pressure is low. Localization nucleates on a plastically weak zone which we have treated as defect in a sample inherited from events prior to the deformation episode under study. 

Our study has shown that the temperature increase due to the conversion of irreversible work (both shear and volumetric) can be an important driver of localization. We observed that the deviatoric (shear) heating outweighs the volumetric contributions, which can be positive or negative, depending on whether expansion or compression are occurring at any given time. Moreover the simulations apear to show that the volumetric deformation facilitates shear deformation, and therefore tends to enhance the heating effect. The presence of dilatancy thus helps to produce a wider shear zone where the volumetric rate of heat generation is stronger. Thermal diffusion also causes the heated zone to spread. The scale of shear bands should therefore depend on both the constitutive law as well as the boundary conditions. 

An important issue relating to the last point concerns how different shear bands may interact. A single shear band may simply tend to localize more and more as a competition develops between local plastic heating and thermal diffusion. On the other hand, two or more differently oriented shear bands may intersect and generate a zone in which localization of deformation is much harder despite the strong heating. Ultimately one shear band may eventually become dominant over the others, but this process will take time, and the temperature evolution of the zone of complex deformation will depend on the geometry and rate of deformation as well as the rheologic law.

In terms of the orientation of the initial seed zone, a critical observation is that the shear bands which develop have a preferred angle, which depends on the frictional, dilatancy, cohesion, deformation history and hardening  parameters \cite{rudnicki1975}. The only difference is that the rate of growth is slowed when the offset between the preferred angle is ${40^\circ}$.   

\subsection{Implications for subduction initiation}

The results of our study hint that large-scale ductile shear bands, which could facilitate subduction intiation, are more likely to nucleate not at the surface, where the rocks are coldest, but at depths where the rocks are hotter than the above threshold temperature as shown in Figure \ref{input_geo_results}. Nevertheless, our results show that lower confining pressures favour shear band formation as long as the rocks are hot enough. Therefore shear bands which initiate at some depth may be able to propagate towards the surface as deformational heating allows isotherms to move upwards to progressively lower pressures around them. 
 
The initial temperature field plays an important role as a result. We have used two initial temperature fields. The obvious implication of the above reasoning is that the steeper the geothermal gradient, the easier it will be for shear bands to initiate at and develop from the base of the lithosphere, because high enough temperatures to facilitate localization by creep will exist at lower confining pressures. Numerical modelling of steady-state subduction utilizes input temperatures which are usually calculated from the cooling of the lithosphere with age, or assuming a pre-existing structure which is usually not known. We have tried to proceed by assuming an initial temperature field that is either laterally uniform or by having two somewhat different geothermal gradients laterally juxtaposed (Figure 16). We observe that the nucleation of deformation on the inherited weak zone can lead to deformed state with a downgoing slab initiatingat the edge of the weak zone. Exploring this idea for intra-oceanic subduction, as observed in a setting like the Lesser Antilles is the aim of future studies.

\section{Conclusion}
We have developed and applied a full 2-D stress update scheme with a consistent algorithmic tangent modulus to handle solid thermomechanical deformation at laboratory and geodynamic scales implementable on the Abaqus finite element solver. Our code handles mechanical and thermal deformations efficiently using 2-D coupled continuum plane strain temperature-displacement triangular elements library (CPE3T) with linear shape functions. For temperature independent deformations, our code is implementable with continuum plane strain elements, axisymmetric elements and plane stress elements depending on the problem at hand.  Theoretical development of the custom constitutive law exploring elasto-thermo-viscoplastic deformation and application to a laboratory-scale and geodynamic problems have afforded us the opportunity to investigate shear band localization and the positive and/or negative feedback to the efficiency of localization and the precursor to the initiation of a subduction zone.

Inlcuding thermal effects, our results show that low temperatures do not necessarily favour the formation of shear bands, but shear bands nucleate at high temperatures and are likley to propagate towards lower pressures. Diffusive processes will also play a role in this propagation but its specific role will depend on the geometry and rate of deformaton imposed on the material domain.

We can also note that the presence of a relatively small weak zone appears to be sufficient to allow shear bands to progressively develop and \enquote{break} the lithosphere without unreasonably high stresses being necessary as the localized deformation propagates into the previously \enquote{stronger} parts of the domain. A further interesting, and perhaps surprising aspect of our results is that when including dilatational deformations and stress components in the computation of the heat source term in the thermal energy equation, we observed that heating can dominate under some conditions because samples underwent dilation, and cooling in few cases. Therefore we may expect to see both heating and cooling zones develop along with the growth of shear bands in some geodynamic situations, presumably depending mainly on the boundary conditions that are  applied. This point deserves more work and analysis beyond beyond the scope of the present paper, and will be presented in a future paper that explores the geodynamic situation more fully. 

\section*{Acknowledgements}
EM and ST would like to thank FEDER
European Community program within
the Interreg Cara\"{i}bes \enquote{PREST} project for funding this work. HSB appreciates the European Research Council grant PERSISMO (grant number 865411) for partial support of this work. 
\section*{Data availability}
All codes developed in the framework of the work reported here are available upon request to the corresponding author.

\end{document}